\documentclass[pdflatex,sn-mathphys-ay]{sn-jnl}

\usepackage{graphicx}%
\usepackage{multirow}%
\usepackage{amsmath,amssymb,amsfonts}%
\usepackage{amsthm}%
\usepackage{mathrsfs}%
\usepackage[title]{appendix}%
\usepackage{xcolor}%
\usepackage{textcomp}%
\usepackage{manyfoot}%
\usepackage{booktabs}%
\usepackage{algorithm}%
\usepackage{algorithmicx}%
\usepackage{algpseudocode}%
\usepackage{listings}%
\usepackage{subfigure}

\usepackage[normalem]{ulem}

\newtheorem{theorem}{Theorem}
\newtheorem{lemma} [theorem] {Lemma}
\newtheorem{corollary} [theorem] {Corollary}
\newcommand{\set}[1]{\left\{#1\right\}}

\newcommand{\brak}[1]{\left<#1\right>}
\newcommand{\abs}[1]{{\left|#1\right|}}
\newcommand{\enset}[2]{\left\{#1 ,\ldots , #2\right\}}

\newcommand{\funcdef}[3]{{#1}:{#2} \to {#3}}

\newcommand{\reals}{{\mathbb{R}}}

\newcommand{\cbcut}{\textsc{CBcut}}
\newcommand{\vcsp}{\textsc{VCSP}}
\newcommand{\wcbcut}{\textsc{WghtCBcut}}
\newcommand{\ecbcut}{\textsc{EdgedCBcut}}

\newcommand{\cbwa}{\cbcut(r, \vw)}

\theoremstyle{definition}

\theoremstyle{definition}

\theoremstyle{definition}

\usepackage{enumitem}

\DeclareMathOperator*{\minimize}{minimize}

\usepackage{mathtools}
\DeclarePairedDelimiter\ceil{\lceil}{\rceil}
\DeclarePairedDelimiter\floor{\lfloor}{\rfloor}
\newcommand{\cut}{\mathbf{cut}}
\newcommand{\st}{$s$-$t$}

\newcommand{\V}{\mathcal{V}}
\newcommand{\E}{\mathcal{E}}

\newcommand{\vw}{\mathbf{w}}
\newcommand{\vW}{\mathbf{W}}

\newcommand{\vhw}{\hat{\mathbf{w}}}

\newcommand{\hw}{\hat{w}}

\usepackage{tikz}

\begin{document}

\title[Article Title]{Improved Hardness and Approximations for Cardinality-Based Minimum $s$-$t$ Cuts Problems in Hypergraphs}


\author[1]{\fnm{Florian} \sur{Adriaens}}\email{florian.adriaens@helsinki.fi}
\author[2]{\fnm{Vedangi} \sur{Bengali}}\email{vedangibengali@tamu.edu}
\author[1]{\fnm{Iiro} \sur{Kumpulainen}}\email{iiro.kumpulainen@helsinki.fi}
\author[1]{\fnm{Nikolaj} \sur{Tatti}}\email{nikolaj.tatti@helsinki.fi}
\author[2]{\fnm{Nate} \sur{Veldt}}\email{nveldt@tamu.edu}

\affil[1]{\orgname{University of Helsinki, HIIT}, \city{Helsinki}, \country{Finland}}
\affil[2]{\orgdiv{Department}, \orgname{Texas A\&M University}, \city{College Station,}, \state{Texas}, \country{USA}}


\abstract{In hypergraphs, an edge that crosses a cut (i.e., a bipartition of nodes) can be split in several ways, depending on how many nodes are placed on each side of the cut. A cardinality-based splitting function assigns a nonnegative cost of $w_i$ for each cut hyperedge $e$ with exactly $i$ nodes on the side of the cut that contains the minority of nodes from $e$. The cardinality-based minimum $s$-$t$ cut aims to find an $s$-$t$ cut with minimum total cost. We answer a recently posed open question by proving that the problem becomes NP-hard outside the submodular region shown by~\cite{veldt2022hypergraph}. Our result also holds for $r$-uniform hypergraphs with $r \geq 4$. Specifically for $4$-uniform hypergraphs we show that the problem is NP-hard for all $w_2 > 2$, and additionally prove that the No-Even-Split problem is NP-hard. We then turn our attention to approximation strategies and approximation hardness results in the non-submodular case. We design a strategy for projecting non-submodular penalties to the submodular region, which we prove gives the optimal approximation among all such projection strategies. We also show that alternative approaches are unlikely to provide improved guarantees, by showing matching approximation hardness bounds assuming the Unique Games Conjecture and asymptotically tight approximation hardness bounds assuming $\text{P} \neq \text{NP}$.}




\maketitle

\section{Introduction}
\label{sec1}
A cut in a graph is a set of edges whose removal partitions the nodes into disconnected components. Finding small graph cuts is a common algorithmic primitive for clustering and partitioning applications such as image segmentation, community detection in social networks, and workload partitioning tasks in parallel computing. In the past several years there has been an increasing interest in solving cut problems over \textit{hypergraphs}, where nodes are organized into multiway relationships called hyperedges~\citep{ccatalyurek2023more, veldt2022hypergraph,zhu2022hypergraph,chekuri2018minimum,panli2017inhomogeneous,panli_submodular}. While edges in a graph model pairwise relationships, hyperedges can directly model multiway relationships such as group social interactions, groups of biological samples with similar gene expression patterns, chemical interactions involving multiple reagents, or groups of interdependent computational tasks in parallel computing applications. 

Given a bipartition of nodes, the standard hypergraph cut function simply counts the number of hyperedges that span both partitions~\citep{lawler1973}. Although this is a straightforward generalization of the graph cut function, it does not capture the fact that there are multiple different ways to split the nodes of a hyperedge into two clusters, each of which may be more or less desirable depending on the application. This has led to recent generalized hypergraph cut functions that assign different cut penalties depending on how the hyperedge is split, with generalized cut penalties captured by a \emph{splitting function} defined for each hyperedge~\citep{veldt2022hypergraph,panli_submodular,panli2017inhomogeneous,zhu2022hypergraph,fountoulakis2021local,chen2023submodular}. Many applications focus specifically on hyperedge cut penalties that are cardinality-based, meaning the penalties depend only on the number of nodes of a hyperedge that are on each side of a split. \cite{veldt2022hypergraph} provided a systematic study of the hypergraph $s$-$t$ cut problem for cardinality-based cut functions, showing that this problem can be reduced to an $s$-$t$ cut problem in a directed weighted graph if and only if all hyperedge splitting functions are submodular. These fundamental primitives for hypergraph $s$-$t$ cuts have since been used as subroutines for other hypergraph analysis techniques, such as localized clustering and semi-supervised learning algorithms on large hypergraphs~\citep{liu2021strongly,veldt2020minimizing}, new approaches for dense subhypergraph discovery~\citep{huang2024densest}, and faster algorithms for decomposable submodular function minimization~\citep{veldt2021approximate}. 

In addition to their algorithmic techniques for submodular cut penalties, \cite{veldt2022hypergraph} proved that the cardinality-based $s$-$t$ cut problem is NP-hard for certain specific non-submodular splitting functions. They also highlighted a trivial setting where penalties are not submodular but an optimal (zero-cost) solution can be found by placing one terminal node ($s$ or $t$) in a cluster by itself. The latter rules out the possibility of showing that the cardinality-based $s$-$t$ cut problem is poly-time solvable if and only if cut penalties are submodular. The complexity of the problem remained unknown for a large class of cut penalty choices. Settling these complexity results, even if only for the case of 4-uniform hypergraphs, was recently included in a list of open problems in applied combinatorics~\citep{aksoy2023seven}. 

\paragraph{The present work: settled hardness and improved approximations}
In our work, we begin by settling the complexity of the cardinality-based $s$-$t$ cut problem and in doing so answer the open questions posed by~\cite{aksoy2023seven}. Concretely, we prove that for every choice of non-submodular weights---except for the trivial case---the cardinality-based hypergraph $s$-$t$ cut problems are NP-hard. We provide two different approaches to show this result. The first is by showing direct reductions from the maximum cut problem, and the second approach draws a connection between generalized hypergraph cut problems and an earlier notion of Valued Constraint Satisfaction Problems (VCSPs) from the theoretical computer science literature~\citep{cohen2006algebraic,cohen2011algebraic}. We can then leverage complexity dichotomy results for the latter problem as another way to prove hardness for non-submodular hypergraph $s$-$t$ cuts. 

Next, we turn to approximation algorithms and approximation hardness results for cardinality-based hypergraph $s$-$t$ problems outside the submodular region. For 4-uniform hypergraphs, the complexity of the problem depends on a single cut penalty $w_2$, that is, the penalty for splitting a hyperedge evenly, and there is a simple strategy for converting a non-submodular splitting function to the closest submodular function, with known approximation factors~\citep{veldt2022hypergraph}.

For larger hyperedges, there can be far more cut penalties since there are more ways to split a hyperedge, and finding the best way to project a non-submodular function to a submodular function is more nuanced. We provide a simple strategy for this projection step by viewing it as a piecewise linear function approximation problem. We prove our strategy provides the optimal approximation factor among all methods that are based on replacing a non-submodular function with a submodular function. We complement these approximation techniques with several approximation hardness results. We first of all establish APX-hardness for problems outside the non-submodular regime. We then leverage the connection to VCSPs to prove that our approximation results based on projecting to the submodular region are the best possible assuming the Unique Games Conjecture.

\section{Technical Preliminaries}
\label{sec:prelims}
We begin by reviewing formal definitions for generalized hypergraph cut problems, along with needed technical background on Valued Constraint Satisfaction Problems.

\subsection{Generalized hypergraph minimum $s$-$t$ cuts}
Consider a hypergraph $\mathcal{H}=(\mathcal{V}, \mathcal{E})$, where each hyperedge $e \in \E$ is a set of $|e|\ge 2$ nodes. Given a set of nodes $S$, a hyperedge is cut when its nodes are split between sets $S$ and $\bar S = V\backslash S$.  
A straightforward extension of the graph cut function to hypergraphs is given by the commonly studied \textit{all-or-nothing} cut function, which simply counts the number of hyperedges crossing a bipartition (or the sum of scalar weights of hyperedges if the hypergraph is weighted)~\citep{lawler1973}. Thus, every way of splitting the nodes of a single hyperedge leads to the same penalty for cutting that hyperedge.

A more general approach involves \textit{hyperedge splitting functions}, which assign a non-negative penalty for each of the $2^e$ potential ways a hyperedge $e$ can be split~\citep{veldt2022hypergraph,li2018submodular,panli2017inhomogeneous}. 
Formally, we define a splitting function $\vw_e:2^e\rightarrow \mathbb{R}$ for each hyperedge $e\in \E$ that satisfies
\begin{alignat*}{2}
	&\vw_e(A) \ge 0, &\quad \quad & \text{for all } A \subseteq e,\\
	&\vw_e(A) = \vw_e(e\setminus A), && \text{for all } A \subseteq e,\\
	&\vw_e(e) = \vw_e(\emptyset) = 0.
\end{alignat*}
For a set $S \subseteq \V$ and $\{s,t\} \subseteq \V$ as source and sink nodes respectively, the generalized hypergraph minimum \st{} cut problem is then defined to be
\begin{align}
	\label{eq:hyperstcut}
	\begin{split}
		&\minimize \quad \textbf{cut}_{\mathcal{H}}(S) = \sum_{e \in \partial S}\vw_e(e \cap S) = \sum_{e \in \E}\vw_e(e \cap S) \quad \text{subject to } \; s\in S, t\in \bar S,
	\end{split}
\end{align}
where $\partial S = \{ e \in \E : e\cap S \neq \emptyset, e\cap \bar{S} \neq \emptyset\}$ is the set of cut hyperedges.

A splitting function is \emph{submodular} if for every $A,B \in 2^e$ it satisfies that
\begin{align}
	\vw_e(A \cap B) + \vw_e(A \cup B) \leq \vw_e(A) + \vw_e(B).
\end{align}
When all splitting functions are submodular, the cut function $\cut_\mathcal{H}$ is a sum of submodular functions and hence submodular. The minimum $s$-$t$ cut problem is then polynomial-time solvable, as it is a special case of submodular function minimization.

\subsection{Cardinality-based minimum \st{} cuts}
\textit{Cardinality-based} functions are defined based only on the number of nodes on each side of the partition. Formally, these functions satisfy the additional condition: 
\begin{align}
	\label{eq:card}
	\vw_e(A) = \vw_e(B) \quad \forall A,B \in 2^e \text{ where } |A|=|B|.
\end{align}
For a hyperedge of size $r = |e|$, $\textbf{w}_e$ can be completely characterized by $q=\floor{r/2}$ splitting penalties denoted as $w_i$ for $i \in \{0,1, \ldots , q\}$, where $\textbf{w}_e(A) = w_i$ is the penalty for every $A \subseteq e$ such that $\min \{|A|,|e\backslash A|\} = i$. Observe that $w_0 = 0$ always. We also often refer to these as \textit{splitting parameters}
as they can be viewed as parameters defining a class of hypergraph $s$-$t$ cut problems. We call a cut edge an $(a,b)$ split if exactly $a$ of its nodes are on one side of the cut and exactly $b$ are on the opposite side. For example, an edge consisting of five nodes is called a $(2,3)$ split, if for a given cut $S$, two of its nodes are part of $S$, while the three others are part of $\bar S$. In this case, the splitting penalty for the cut edge is $w_2$. We remark that in practice, each different $e \in E$ may be associated with a different set of splitting penalties.

\paragraph{The $\cbwa$ problem}
In practice, hyperedges can be of different sizes and there may be reasons to consider associating different splitting functions to different hyperedges. However, for the purpose of understanding fundamental tractability results, we restrict our attention to $r$-uniform hypergraphs where all hyperedges have the same cardinality-based splitting function. Tractability and hardness results for other cardinality-based hypergraph \st{} cut problems (where hyperedges can have different sizes and splitting functions) can be easily determined by extending results for the uniform case.

Formally, let $\mathcal{H} = (\V, \mathcal{E})$ be an $r$-uniform hypergraph and let $\vw = (w_1, w_2, \hdots, w_q)$ be a set of non-negative splitting penalties where $q = \floor{r/2}$. We refer to $\vw$ as the splitting vector. The $\cbwa$ problem is given by
\begin{align}
	\label{eq:cbcut}
	\minimize \quad \textbf{cut}_{\vw}(S) = \sum_{i=1}^{q} w_i \cdot |\partial S_i|, \quad \text{ subject to } s \in S \text{ and } t \in \bar{S},
\end{align}
where $\partial S_i = \{e \in \E:|S\cap e| \in \{i,r-i\}\}$ is the set of $(i,r-i)$ split hyperedges.
For some of our results, we also consider a weighted variant of the problem, where we assume that each hyperedge $e \in \E$ is associated with a scalar rational cost $c_e \geq 0$, and we scale the cut penalty at this edge by this weight. Formally, the $\wcbcut(r, \vw)$ problem is then defined to be
\begin{align}
	\label{eq:weightedcard}
	\minimize \quad \textbf{cut}_{\vw}(S) = \sum_{i=1}^{q} w_i \cdot C(\partial S_i), \quad \text{ subject to } s \in S \text{ and } t \in \bar{S},
\end{align}
where $C(\partial S_i) = \sum_{e \in \partial S_i} c_e$. 

We typically treat the splitting values $(w_1, w_2, \hdots, w_q)$ as fixed constants. However, in some cases it is also interesting to consider a variant where $w_i = \infty$ for some integer $i$, representing a hard constraint that prohibits any $(i, r-i)$ splits. One example is the previously considered \textsc{No-even-split} cut problem (which can be denoted 
$\cbcut(4,(w_1 = 1, w_2 = \infty)$), where the goal is to find a minimum $s$-$t$ cut of a 4-uniform hypergraph while strictly prohibiting any $(2,2)$-splits. Settling the complexity of this problem was included in the list of open questions in applied combinatorics of~\cite{aksoy2023seven}.

\paragraph{Prior tractability and hardness results}
\cite{veldt2022hypergraph} proved that a cardinality-based splitting function for an $r$-node hyperedge is submodular if and only if its splitting penalties satisfy
\begin{align}
	2w_1 &\geq  w_2, \nonumber\\
	2w_j &\geq  w_{j-1}+w_{j+1}, & & \text{ for } j= 2,\hdots,q-1,\label{eq:submodular}\\
	w_q &\geq  w_{q-1} \ge \hdots \ge w_2 \ge w_1 \ge 0. \nonumber
\end{align}
Thus, if $\vw = (w_1, w_2, \hdots, w_q)$ satisfies these conditions, $\cbwa{}$ is polynomial-time solvable. The problem is also (trivially) polynomial-time solvable if $w_1 = 0$ since in this case one can set $S = \{s\}$ and the resulting cut penalty is zero. This holds independent of the values for $(w_2, w_3, \hdots, w_q)$, including values for which the overall splitting function is non-submodular. 
If $w_1 > 0$, we can scale all penalties without loss of generality so that $w_1 = 1$; we will typically assume this scaling for all cases we consider throughout the manuscript. \cite{veldt2022hypergraph} proved that $\cbwa$ is NP-hard, via reduction from maximum cut, for splitting penalties in the non-submodular regime for which $0 < \min(\vw) < w_1$.

\subsection{Valued Constraint Satisfaction Problems}
\textit{Valued Constraint Satisfaction Problems} (VCSPs) provide a general framework to model and solve optimization problems over a language involving variables, constraints, and value assignments. \cite{cohen2004complete} studied the complexity of VCSPs defined over Boolean variables. We will model the $\cbwa$ problem as a special type of Boolean VCSP and then translate existing complexity results for the latter problem to hypergraph cut problems. We review the notation and definitions established by~\cite{cohen2004complete}.\footnote{These authors gave generalized definitions for VSCPs that also apply to the non-Boolean case, but we restrict to Boolean variables since this suffices for our purposes.}

An instance of Boolean VCSP is given by a tuple $\mathcal{P} = \brak{V,\mathcal{X}, \Theta}$, where
\begin{itemize}
	\item $V = \{v_1, v_2, \hdots, v_n\}$ is a finite set of Boolean variables,
	\item $\mathcal{X}\subseteq \reals^+$ is a set of possible costs,
	\item $\Theta$ is a set of constraints, each defined by a pair $\theta = \brak{\sigma,\phi}$, where the \emph{scope} $\sigma \subseteq V$ defines a set of variables the constraint applies to, and $\funcdef{\phi}{\{0,1\}^{|\sigma|}}{\mathcal{X}}$ is a \emph{cost} function that maps every possible assignment of these variables to a cost in $\mathcal{X}$.
\end{itemize}

Given a function $\funcdef{\phi}{\{0,1\}^m}{\mathcal{X}}$, the value $m$ is called the \emph{arity} of $\phi$, and it only applies to scopes of size $m$.

Given a scope $\sigma \subseteq V$, we will write $\sigma_i = j$ if $v_j$ is the $i$-th element in $\sigma$. For example, if $\sigma = \{v_2, v_4, v_{10}\}$, then $|\sigma| = 3$, $\sigma_1 = 2$, $\sigma_2 = 4$, and $\sigma_{3} = 10$. The goal is to find an assignment $a \colon V \rightarrow \{0,1\}^n$ of variables to Boolean values to solve the following problem,
\begin{equation}
	\label{eq:vcsp}
	\minimize_{a \colon V \rightarrow \{0,1\}^n} \quad \textbf{cost}_{\mathcal{P}}(a) = \sum_{\langle \sigma,\phi\rangle \in \Theta} \phi(a(v_{\sigma_1}),a(v_{\sigma_2}),\hdots,a(v_{\sigma_{|\sigma|}})).
\end{equation}
Prior work has focused on proving complexity results for this objective under different assumptions about the cost functions $\phi$. Formally, let $\Gamma$ represent a collection of cost functions. A \textit{Valued Boolean Constraint Language} $\vcsp(\Gamma)$  is then a tuple $\brak{V, \mathcal{X}, \Theta, \Gamma}$ where cost functions in the constraint set $\Theta$ come from the collection $\Gamma$. The constraint language is called \emph{tractable} when all instances within $\vcsp(\Gamma)$ can be solved in polynomial time. It is {NP-hard} (as an entire language) if an existing NP-hard problem has a polynomial-time reduction to $\vcsp(\Gamma)$.

\paragraph{Relation to hypergraph $s$-$t$ cuts} There is a close connection between the VCSP objective in~\eqref{eq:vcsp} and the generalized hypergraph $s$-$t$ cut problem in~\eqref{eq:hyperstcut}. The Boolean variables $V$ can be thought of as (non-terminal) nodes $\V - \{s,t\}$ in a hypergraph $\mathcal{H} = (\V, \E)$. Assigning a variable to $1$ corresponds to placing a node on the $s$-side of a cut, while assigning to $0$ means placing the node on the $t$-side. Each scope $\sigma$ of a constraint in the VCSP problem corresponds to the nodes in some hyperedge $e$, and the cost function $\phi$ corresponds to the splitting function $\vw_e$, which gives a penalty for each way of splitting the nodes. Proving tractability results for a cost function collection $\Gamma$ then corresponds to proving tractability results for a collection of hypergraph $s$-$t$ cut problems defined by a class of splitting functions (in our case, cardinality-based splitting functions for a specific choice of splitting penalties $\vw$). This relationship allows us to translate existing tractability results from the VCSP literature to hypergraph cut problems, though a few additional details are needed to ensure the relationship is formalized correctly (see Section~\ref{sec:nphard}).

\paragraph{Complexity dichotomy results for Boolean VCSPs}
\cite{cohen2004complete} proved complete complexity dichotomy results for Boolean VCSPs, which rely on proving certain inequalities involving the notion of a \emph{multimorphism}. To summarize these results, we must first extend a cost function $\phi \colon \{0,1\}^m \rightarrow \mathcal{X}$ of arity $m$ so that it can be applied to $m$ \emph{tuples} of Boolean values, rather than just $m$ Boolean values. Formally, if $\{t_1, t_2, \hdots, t_m\} \subseteq \{0,1\}^k$ is a set of $m$ Boolean vectors of size $k$, where $t_j[i]$ is the $i$-th entry of the $j$-th vector, we define
\begin{equation}
	\phi(t_1,t_2,\ldots,t_m) = \sum_{i=1}^k \phi(t_1[i],t_2[i],\ldots,t_m[i]).
\end{equation}
In other words, evaluating $\phi$ on these tuples corresponds to evaluating them $k$ times (one for each position in the vectors), and then summing the results. A function $F \colon \{0,1\}^k \rightarrow \{0,1\}^k$ is defined to be a \emph{multimorphism} of $\phi$ if the following inequality holds:
\begin{equation}
	\label{eq:multimorphism}
	\phi(F(t_1),F(t_2),\ldots,F(t_m)) \le \phi(t_1,t_2,\ldots, t_m).
\end{equation}
If $\Gamma$ is a collection of cost functions, then we say $F$ is a multimorphism of $\Gamma$ if $F$ is a multimorphism of every $\phi \in \Gamma$. \cite{cohen2004complete} proved the following complexity dichotomy result for languages with finite-valued cost functions.

\begin{theorem}[Corollary 2~\citep{cohen2004complete}]
	\label{thm:vcsp-trac}
	Assume an instance $\vcsp(\Gamma) = \brak{V, \mathcal{X}, \Theta, \Gamma}$.
	Assume that $\mathcal{X}$ has only finite values. Define $\funcdef{F_0}{\set{0, 1}}{0}$, $\funcdef{F_1}{\set{0, 1}}{1}$, and
	$\funcdef{F_2}{\set{0, 1}^2}{\set{0, 1}^2}$ as
	\[
	F_0(t) = 0,\quad
	F_1(t) = 1,\quad
	F_2(t_1, t_2) = (\min \set{t_1, t_2}, \max \set{t_1, t_2}).
	\]
	Then
	$\vcsp(\Gamma)$ is tractable if $F_0$ or $F_1$ or $F_2$ are multimorphisms of $\Gamma$; otherwise $\vcsp(\Gamma)$ is NP-hard. 
\end{theorem}
It is known that $F_2$ is a multimorphism of a cost function $\phi$ if and only if $\phi$ is submodular~\citep{cohen2003soft,cohen2004complete}. We provide a proof to give an intuition for multimorphisms.

\begin{lemma}
	\label{lem:submodular}
	The cost function $\phi \colon \{0,1 \}^m \rightarrow \reals^+$ is submodular if and only if it has $F_2$ as a multimorphism.
\end{lemma}
\begin{proof}
	The definition of submodularity given for set functions in Eq.~\eqref{eq:submodular} can be translated easily to a property of a Boolean function $\phi$. Formally, consider two $m$-tuples of binary values $\textbf{a} = (a_1, a_2, \hdots, a_m)$ and $\textbf{b} = (b_1, b_2, \hdots, b_m)$, which we can think of as indicator vectors for some sets $A$ and $B$. The indicator vectors for sets $A \cup B$ and $A \cap B$ are given by:
	\begin{align}
		\textbf{a}\cap\textbf{b} &= (\min \{a_1, b_1\}, \min \{a_2, b_2\}, \hdots ,\min \{a_m, b_m\})\\
		\textbf{a}\cup\textbf{b} &= (\max \{a_1, b_1\}, \max \{a_2, b_2\}, \hdots ,\max \{a_m, b_m\}).
	\end{align}
	By definition, $\phi$ is submodular if it satisfies the constraint
	\begin{equation} 
		\label{eq:booleansubmodularity}
		\phi(\textbf{a}\cap\textbf{b}) + \phi(\textbf{a}\cup\textbf{b}) \leq \phi(\textbf{a}) + \phi(\textbf{b})
	\end{equation}
	for an arbitrary pair of Boolean vectors \textbf{a} and \textbf{b}. 
	Define now a set of $m$ 2-tuples $\{t_1, t_2, \hdots, t_m\}$ by stacking \textbf{a} and \textbf{b}, so that $t_i = (\textbf{a}[i], \textbf{b}[i])$. The definition of a multimorphism in Eq.~\eqref{eq:multimorphism} applied to $F_2$ exactly corresponds to the inequality defining submodularity in Eq.~\eqref{eq:booleansubmodularity}.
\end{proof}

\section{NP-hardness for Non-submodular Regime}
\label{sec:nphard}
We now prove NP-hardness for all cardinality-based $s$-$t$ cut problems with non-submodular parameters and $w_1 > 0$. Recall that when $w_1 = 0$ the problem is trivial. We provide two different approaches for showing NP-hardness. The first is a direct reduction from the NP-hard \textsc{MaxCut} problem. 
Our second approach draws a connection to Valued Boolean Constraint Satisfaction Problems to show how existing hardness results for this problem prove corresponding hardness for cardinality-based $s$-$t$ cuts. The latter approach to proving hardness is less direct than our direct reductions from \textsc{MaxCut}, but makes a useful connection to another related body of research that is useful for establishing later approximation hardness results.

The main conclusion of this section is to confirm that (except for the trivial case $w_1 = 0$) submodularity not only coincides exactly with graph-reducibility (as shown by~\cite{veldt2022hypergraph}), but also coincides with tractability.
\begin{theorem}
	\label{thm:nphard}
	If $w_1 = 1$, $\cbwa$ is tractable if and only if the splitting function is submodular. 
\end{theorem}

\subsection{Direct reductions from \textsc{MaxCut}}
\label{sec:maxcut_reductions}

\begin{proof}[Direct proof of Theorem~\ref{thm:nphard}]
	We prove the claim in two cases, depending on which submodular constraint from Eq. (\ref{eq:submodular}) is violated. For simplicity, we will write $w_0 = 0$ to merge the first two constraints into one case.
	
	\textbf{The case $w_{i-1} + w_{i+1} > 2w_i$  for some $1 \leq i \leq q-1$.}
	Let $G=(V,E)$ be an instance graph of \textsc{MaxCut} with $m$ edges.
	
	Let $\alpha = \max(\lceil m(w_{i-1} + w_{i+1}) + 1 \rceil, 2r, 10)$.
	Construct a hypergraph $\mathcal{H}$ as follows. 
	First, add two sets of nodes, $S = \{s_1,\ldots,s_{\alpha}\}$ and $T = \{t_1,\ldots,t_{\alpha}\}$. Let $s_1$ be the source and $t_1$ the sink of our cut problem.
	We connect every set of $r$ nodes in $S$ with a hyperedge. We also repeat the same step for $T$.
	
	For every edge $(x,y) \in E$, add two hyperedges 
	$(s_1,\ldots,s_{i-1},x,y,t_1,\ldots,t_{r - i - 1})$ and 
	$(t_1,\ldots,t_{i-1},x,y,s_1,\ldots,s_{r - i - 1})$ to $\mathcal{H}$.

	Consider any cut which partitions $S$ into two pieces $S_1$ and $S_2$ where $|S_1| \geq |S_2| \geq 1$.
	Since $\alpha \geq 2r$ and $\alpha \geq 10$, we have $\abs{S_1} \geq r$ and $\abs{S_1} \geq 5$ so ${\abs{S_1} \choose r - 1} \geq {\abs{S_1} \choose 2} \geq 2 \abs{S_1}$. We then have that there are 
	\[{\abs{S_1} \choose r - 1}{\abs{S_2}} + {\abs{S_2} \choose r - 1}{\abs{S_1}} \geq {\abs{S_1} \choose 2} \geq 2 \abs{S_1} \geq \alpha \] 
	hyperedges that are $(r - 1, 1)$ split, yielding a cost of at least $\alpha w_1 = \alpha$. A similar conclusion holds for any cut that separates the nodes of $T$.
	
	Keeping $S$ on one side, and the remaining nodes on the other side, results in a cut of cost $m w_{i - 1} + m w_{i + 1} < \alpha$.
	Consequently, $S$ and $T$ are not cut in the optimal cut.
	
	If $x$ and $y$  are on different sides of the cut, the two hyperedges containing these nodes are split with a combined cost of $2w_i$. If $x$ and $y$ are on the same side of a cut, one hyperedge is cut with a cost of $w_{i-1}$ and the other with a cost of $w_{i+1}$.
	If we cut $c$ edges in the instance of \textsc{MaxCut}, total cost of the corresponding cut in $\mathcal{H}$ is
	\begin{equation*}
		\text{cost} = (m-c)(w_{i-1} + w_{i+1})+2cw_i = m(w_{i-1} + w_{i+1}) + (2w_i-(w_{i-1} + w_{i+1}))c.
	\end{equation*}
	
	If $w_{i-1} + w_{i+1} > 2w_i$, then minimizing the cost of the cut in $\mathcal{H}$ corresponds to finding a \textsc{MaxCut} in $G$.
	
	\textbf{The case $w_i>w_{i+1}$ for some $1 \leq i \leq q-1$.}
	Let $\alpha = \max(\lceil mw_i + 1 \rceil, 2r, 10)$. Create $S$ and $T$ as before with $s_1$ the source and $t_1$ the sink of our cut problem. We also add a set of $\alpha$ vertices $U_x$ for each vertex $x$, and introduce a hyperedge for each $r$-tuple of nodes in $U_x$.
	
	For every $(x,y) \in E$, add one hyperedge $(s_1,\ldots,s_i,u_1, \ldots, u_{r - 2i - 1},y_1,t_1,\ldots,t_{i})$ to $\mathcal{H}$, where $u_j \in U_x$ and $y_1 \in U_y$. We see similarly that cutting $S$, $T$, or $U_x$ would result in a cost of at least $\alpha$, which is more expensive than a cut isolating $S$ with a cost of $m w_i < \alpha$. Therefore, $S$, $T$, and $U_x$ cannot be cut in the optimal solution.
	
	If $x$ and $y$ are on different sides of the cut, the cost of cutting the associated hyperedge is $w_{i+1}$. If $x$ and $y$ are on the same side of a cut, the cost of cutting is $w_{i}$. 
	If we cut $c$ edges in the \textsc{MaxCut} instance, the total cost of the corresponding cut in $\mathcal{H}$ is
	\begin{equation*}
		\text{cost} = (m-c)w_i + cw_{i+1} = mw_i + (w_{i+1}-w_i)c.
	\end{equation*}
	
	If $w_i > w_{i+1}$, then minimizing the cost of the cut in $\mathcal{H}$ corresponds to finding a \textsc{MaxCut} in $G$.
\end{proof}

\subsection{Hardness via VCSPs}
\label{sec:vcsp}
We now introduce a specific Valued Boolean Constraint Language that we will show exactly corresponds to $\cbwa$, and use this connection as another way to prove Theorem~\ref{thm:nphard}. 
For an integer $r$ and splitting vector $\vw$, we define $\Gamma_{r,\vw}$ to be the constraint language that includes exactly four cost functions $\{\phi_r,\phi_s,\phi_t,\phi_{st}\}$.
We first define the cost function $\phi_r \colon \{0,1\}^r \rightarrow \enset{w_1}{w_q}$ by 
\begin{align}
	\phi_r(x_1,x_2,\hdots, x_r) = w_{\min(j, r - j)},
	\quad\text{where}\quad
	j = \sum_{i = 1}^r x_i.
\end{align}
This corresponds to a cardinality-based splitting function with parameters $\vw$, applied to a hyperedge with $r$ non-terminal nodes. The other three cost functions $\{\phi_s, \phi_t, \phi_{st}\}$ correspond to the same splitting function applied to hyperedges containing $s$ but not $t$, containing $t$ but not $s$, or containing both $s$ and $t$, respectively. We think of $x_i = 1$ as assigning a variable to the $s$-side of the cut and $x_i = 0$ as assigning a variable to the $t$-side. These three cost functions are therefore defined by applying $\phi_r$ with one or two input variables fixed to 0 or 1,
\begin{align*}
\phi_s(x_1, x_2, \hdots, x_{r-1}) &= \phi_r(1,x_1, x_2, \hdots, x_{r-1}), \\
\phi_t(x_1, x_2, \hdots, x_{r-1}) &= \phi_r(x_1, x_2, \hdots, x_{r-1}, 0), \\
\phi_{st}(x_1, x_2, \hdots, x_{r-2}) &= \phi_r(1,x_1, x_2, \hdots, x_{r-2}, 0).
\end{align*}

\begin{lemma}
\label{lem:equivalence}
$\vcsp(\Gamma_{r, \vw})$ is tractable if and only if $\cbwa$ is tractable.
\end{lemma}
\begin{proof}
The reduction in both directions is straightforward; we show one direction for clarity. Let $\mathcal{H} = (U,\E)$ denote an $r$-uniform hypergraph with node set $U = \{u_0 = s, u_1, u_2, \hdots, u_n, u_{n+1} = t\}$. For a hyperedge $e$, let $e(i)$ denote the index of the $i$th node in $e$, so that the hyperedge can be expressed as $e = (u_{e(1)}, u_{e(2)}, \hdots, u_{e(r)}) \subseteq U$. If $e$ contains $s$, we assume nodes in $e$ are ordered so that $u_{e(1)} = s$. If $e$ contains $t$, we assume $u_{e(r)} = t$. The ordering for non-terminal nodes is arbitrary.  

To reduce this instance of $\cbwa$ to an instance of $\vcsp(\Gamma_{r,\vw})$, we introduce a set of $n$ variables $V = \{v_1, v_2, \hdots, v_n\}$. For a hyperedge $e$ not containing terminal nodes, we add a constraint $\langle \langle v_{e(1)}, v_{e(2)}, \hdots, v_{e(r)} \rangle, \phi_r \rangle$. If $e$ contains $s$ but not $t$, we add constraint $\langle \langle v_{e(2)}, \hdots, v_{e(r)} \rangle, \phi_s \rangle$ (since $u_{e(1)} = s$). If it contains $t$ but not $s$, we add constraint $\langle \langle v_{e(1)}, v_{e(2)}, \hdots, v_{e(r-1)} \rangle, \phi_t \rangle$ (since $u_{e(r)} = t$). If it contains both $s$ and $t$, we add constraint $\langle \langle v_{e(2)}, v_{e(3)}, \hdots, v_{e(r-1)} \rangle, \phi_{st} \rangle$. It is straightforward to check that there is a variable assignment with cost $\alpha \geq 0$ for the VCSP instance if and only if there is an $s$-$t$ cut with cut value $\alpha$ in the hypergraph $\mathcal{H}$. The reduction from $\vcsp(\Gamma_{r,\vw})$ to $\cbwa$ is similar. 
\end{proof}

At this point we simply need to interpret Theorem~\ref{thm:vcsp-trac} to determine complexity dichotomy results from the constraint language $\Gamma_{r,\vw}$, which in turn gives complexity dichotomy results for $\cbwa$ that amount to another proof of Theorem~\ref{thm:nphard}. 

\begin{proof}[Proof of Theorem~\ref{thm:nphard} via VSCP equivalence]
Note that all of the cost functions $\{\phi_r, \phi_s, \phi_t, \phi_{st}\}$ are submodular if and only if $\phi_r$ is submodular, which is true if and only if splitting parameters $\vw$ satisfy the submodularity inequalities in Eq.~\eqref{eq:submodular}. We know from Lemma~\ref{lem:submodular} that $F_2$ is a multimorphism of $\phi_r$ if and only if $\phi_r$ is submodular. This corresponds to the known tractable submodular regime for $\cbwa$. 

Theorem~\ref{thm:vcsp-trac} tells us that the only other situation where VCSP($\Gamma_{r,\vw}$) is tractable is when $F_0$ or $F_1$ is a multimorphism of $\Gamma_{r,\vw}$.
Observe that $F = F_0$ is a multimorphism of $\Gamma_{r,\vw}$ if and only if all four of the following inequalities hold for all choices of $\{x_1, x_2, \hdots, x_r\}$,
\begin{align*}
	\phi_r(F(x_1),\hdots,F(x_r)) &= \phi_r(0,\hdots,0)
	= 0 \le \phi_r(x_1,\hdots,x_r), \\
	\phi_s(F(x_1),\hdots,F(x_{r-1})) & = \phi_s(0,\hdots,0)
	= \phi_r(1, 0,\hdots,0) = w_1 \le \phi_s(x_1,\hdots,x_{r-1}),\\
	\phi_t(F(x_1),\hdots,F(x_{r-1})) & = \phi_t(0,\hdots,0)
	= \phi_r(0,\hdots,0) = 0 \le \phi_t(x_1,\hdots,x_{r-1}), \\
	\phi_{st}(F(x_1),\hdots,F(x_{r-2})) & = \phi_{st}(0,\hdots,0)
	= \phi_r(1, 0,\hdots,0) = w_1 \le \phi_{st}(x_1,\hdots,x_{r-2}).
\end{align*}
The first and third inequalities are always true, but the second and fourth are true for all inputs if and only if $w_1 = 0.$ We can similarly show that $F_1$ is a multimorphism of $\Gamma_{r,\vw}$ if and only if $w_1 = 0$. Thus, aside from the submodular regime, the only tractable case for $\cbwa$ is when $w_1 = 0$.
\end{proof}

\section{Approximating Non-submodular Cut Problems}
\label{sec:projections}
Having established complexity dichotomy results for $\cbwa$, we would like to determine the best approximation guarantees we can achieve for the NP-hard non-submodular cases. We will specifically design approximation algorithms that rely on projecting a set of non-submodular splitting penalties (i.e., values $(w_1 = 1, w_2, \hdots, w_q)$ that \emph{do not} satisfy inequalities in~\ref{eq:submodular}) to a nearby set of submodular penalties (``nearby'' values $(\hat{w}_1, \hat{w}_2, \hdots, \hat{w}_q)$  that do satisfy these inequalities). We can then solve the latter submodular hypergraph $s$-$t$ cut problem to provide an approximation for the original non-submodular problem. 

\begin{figure}
	\centering
	\subfigure[$r = 4 \text{ or } 5$]{%
		\centering
		\includegraphics[width=0.35\linewidth]{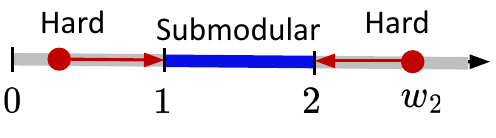}
		\label{fig:4cb-projection}
	}
	\hfill
	\subfigure[$r = 6 \text{ or } 7$]{%
		\centering
		\includegraphics[width=0.35\linewidth]{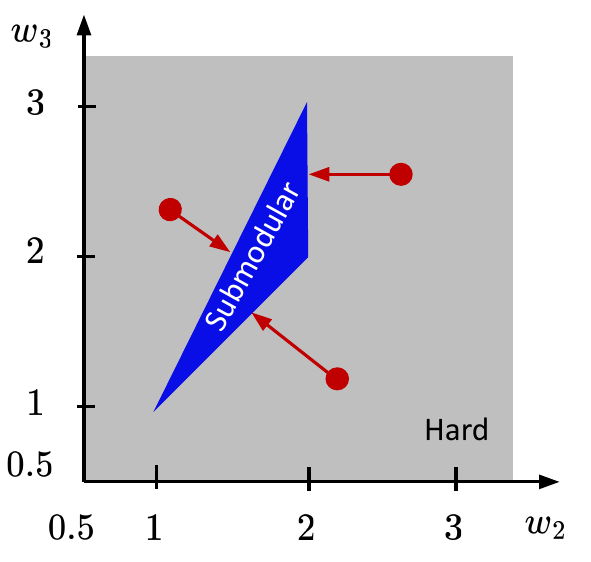} %
		\label{fig:67cb-projection}
	}
	\caption{Submodular (shaded in blue) and NP-hard (shaded in gray) regions of $\cbwa$ for different values of $r$ and  fixed $w_1 = 1$.}
	\label{fig:}
\end{figure}

\paragraph{Known approximations for $\cbcut(4, \vw)$} ~\cite{veldt2022hypergraph} previously applied this approach to obtain simple approximation guarantees for $\cbwa$ for $r \in \{4,5\}$ in the non-submodular region, i.e., when $w_1 = 1$ and $w_2 \notin [1,2]$. Finding a cardinality-based splitting function that is ``closest'' to a non-submodular splitting function is particularly easy in this case since it just involves either decreasing or increasing $w_2$, depending on whether it lies to the right or left of the submodular region $[1,2]$; see Figure~\ref{fig:4cb-projection}. To approximate the problem when $w_2 < 1$, one can compute the solution to the $s$-$t$ cut problem with splitting penalty $\hat{w}_2 = 1$. It is not hard to show this produces a $\frac{1}{w_2}$-approximate solution for the original NP-hard problem. Similarly, if $w_2 > 2$, one can solve the submodular problem where $\hat{w}_2 = 2$ to obtain a $w_2/2$ approximation.

\paragraph{Projecting splitting penalties for $r>5$}
Projecting non-submodular splitting penalties to the submodular region becomes more nuanced when the hyperedge size is $r > 5$; see Figure~\ref{fig:67cb-projection}. This typically involves changing multiple splitting penalties (not just $w_2$), and it is not immediately clear what it means to find the ``best'' or ``closest'' set of submodular splitting penalties for a given splitting vector $\vw$. The submodularity inequalities in Eq.~\eqref{eq:submodular} define a closed convex set, so a natural idea is to project a given set of non-submodular penalties $(w_1 = 1, w_2, \hdots, w_q)$ to this convex set in a way that minimizes a 2-norm distance or the distance in terms of some other norm. There are many well-known methods for projecting a point onto a convex set that could be used for this approach. However, will this provide the best possible approximation guarantee for a non-submodular hypergraph $s$-$t$ cut problem? To answer this question, we will first show how to characterize the approximation guarantee for any technique that projects non-submodular penalties to the submodular region. We will then show how to obtain the best approximation guarantee among all such projection methods by solving a piecewise linear function approximation problem.

\subsection{Approximation bounds for non-submodular cuts}
The space $\mathbb{R}_+^q$ represents the universe of all potential $\cbwa$ splitting vectors. 
Within this space, let $\mathcal{S}_q \subseteq \mathbb{R}_+^q $ represent the subset of vectors that correspond to submodular cardinality-based splitting functions. 
For a vector $\vw \notin \mathcal{S}_q$, we wish to design a method that projects $\vw$ onto a vector $\hat{\vw} = (\hat{w}_1,\hat{w}_2,\hdots,\hat{w}_q) \in \mathcal{S}_q$
in a way that provides the best approximation guarantee for $\cbcut(r, \vw)$. Let $\cut_\vw$ denote the cut function for non-submodular splitting vector $\vw$ and ${\cut}_{\hat{\vw}}$ be the cut function for $\hat{\vw}$. Formally, our goal is to choose $\hat{\vw}$ so that for every $r$-uniform hypergraph $\mathcal{H} = (\V, \E)$ and every $S \subseteq \V$ we have
\begin{align}
	\cut_{\vw}(S) \le {\textbf{cut}}_{\hat{\vw}}(S) \le \rho\cdot\cut_{\vw}(S) \label{eq:max_approx}
\end{align}
for the smallest possible value of $\rho$. Satisfying the first inequality $\cut_{\vw}(S) \le {\textbf{cut}}_{\hat{\vw}}(S)$ amounts to the constraint $\hat{\vw} \ge \vw$. We can enforce this without loss of generality; if the projected vector had a parameter $\hat{w}_j < w_j$ for some $j \in \{1,2,\hdots,q\}$, we could scale the entire vector $\hat{\vw}$  by a factor of $w_j/\hat{w}_j$ without affecting optimal solutions or approximation guarantees. For convenience, we assume throughout this section that $\vw(1) = w_1 = 1$ for the original vector $\vw$ that we wish to project to the submodular region. Note that because of the constraint $\hat{\vw} \geq \vw$, it is possible for $\hat{w}_1 > 1$ to hold. 

The following lemma shows how well we can approximate $\cbwa$ for a splitting vector $\vw$ by solving a nearby problem defined by vector $\hat{\vw}$. The result holds for an arbitrary pair of splitting vectors satisfying $\vw \leq \hat{\vw}$. In principle, the idea is to apply this to project a vector $\vw \notin \mathcal{S}_q$ to a nearby vector $\hat{\vw} \in \mathcal{S}_q$.
\begin{lemma}
	\label{lem:projbound}
	Let $\vw, \hat{\vw} \in \mathbb{R}^q_+$ be a pair of splitting vectors satisfying $\vw \leq \hat{\vw}$. For every $r$-uniform hypergraph $\mathcal{H}=(\V,\E)$ and $S \subseteq \V$, the inequalities in~\eqref{eq:max_approx} are satisfied for
	\begin{align*}
		\rho = \max_i \frac{\hat{w}_i}{w_i}  \quad  \text{ for }  i \in \{1,2,3,\hdots,q\}.
	\end{align*}
\end{lemma}
\begin{proof}
	For an arbitrary set $S \subseteq \V$, the cut penalty with respect to $\vw$ is $\textbf{cut}_{\vw}(S)$ and the cut value with respect to $\hat{\vw}$ is ${\textbf{cut}}_{\hat{\vw}}$. Thus:
	\begin{align*}
		\frac{\textbf{cut}_{\hat{\vw}}(S)}{\textbf{cut}_{\vw}(S)}
		&= \frac{\hat{w}_1\cdot|\partial S_1| + \hat{w}_2\cdot |\partial S_2| + \cdots + \hat{w}_q\cdot |\partial S_q|}{w_1\cdot |\partial S_1| + w_2\cdot|\partial S_2| + \cdots + w_q\cdot |\partial S_q| } \le \max_{i \in \{1,\hdots, q\}} \left( \frac{\hat{w}_i}{w_i} \right).
	\end{align*}
\end{proof}
Observe that the approximation bound of $\max_i \frac{\hat{w}_i}{w_i}$ is tight. Recall that our goal is to project non-submodular splitting penalties in such a way that we can guarantee a certain approximation factor for all instances of $\cbwa$. This means that we need the inequality in~\eqref{eq:max_approx} to hold for every possible $r$-uniform hypergraph $\mathcal{H}$ and node set $S$. In the worst case, it is possible to construct a hypergraph such that the approximation factor in Lemma~\ref{lem:projbound} is tight. More precisely, if the largest ratio is $\hat{w}_t/w_t$ for some $t \in \{1, 2, \hdots, q \}$, we can find a hypergraph with node set $S$ where $\partial S$ only contains cut hyperedges with exactly $t$ nodes on the small side of the cut. This results in an approximation factor of exactly $\hat{w}_t/w_t$.

\subsection{Norm-minimizing projection techniques}
\begin{figure}[t]
	\centering
	\subfigure[Bounds obtained via $\ell_1$-norm projection]{%
		\includegraphics[width=0.45\textwidth]{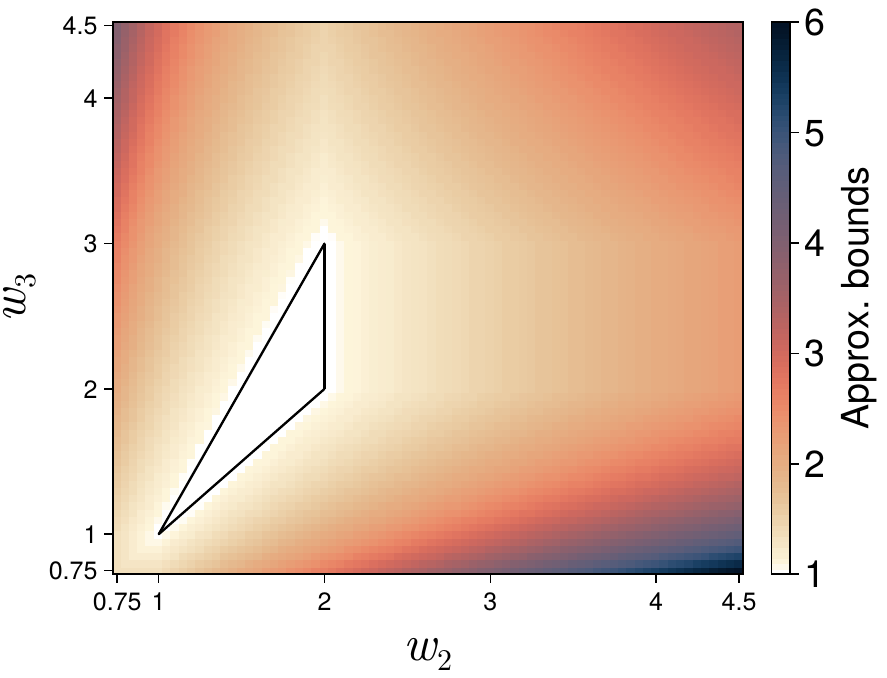}
		\label{fig:heatmap_L1}
	}
	\hfill
	\subfigure[Bounds obtained via $\ell_2$-norm projection]{%
		\includegraphics[width=0.45\textwidth]{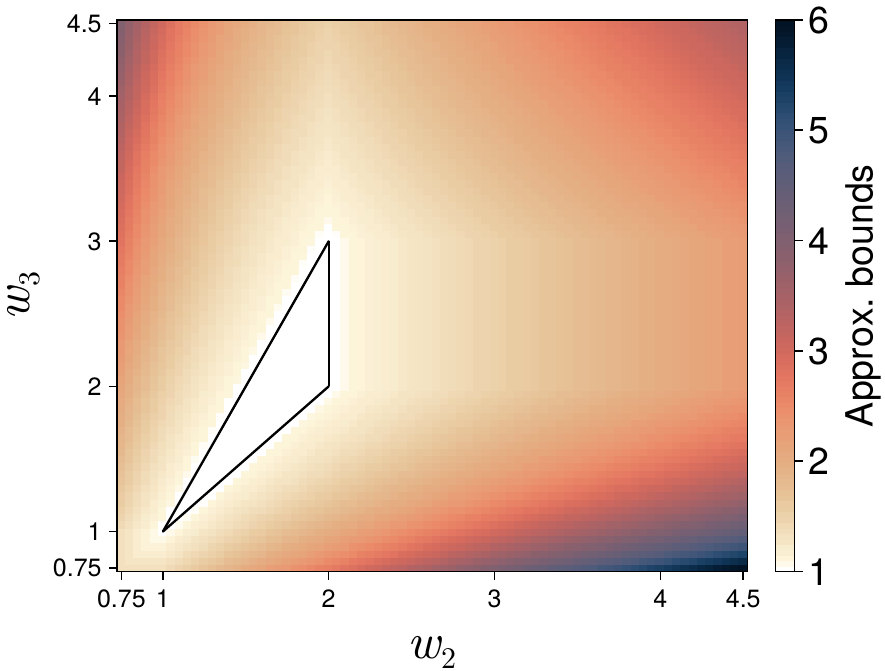} %
		\label{fig:heatmap_L2}
	}
	\hfill
	\subfigure[Bounds obtained via $\ell_\infty$-norm projection]{%
		\includegraphics[width=0.45\textwidth]{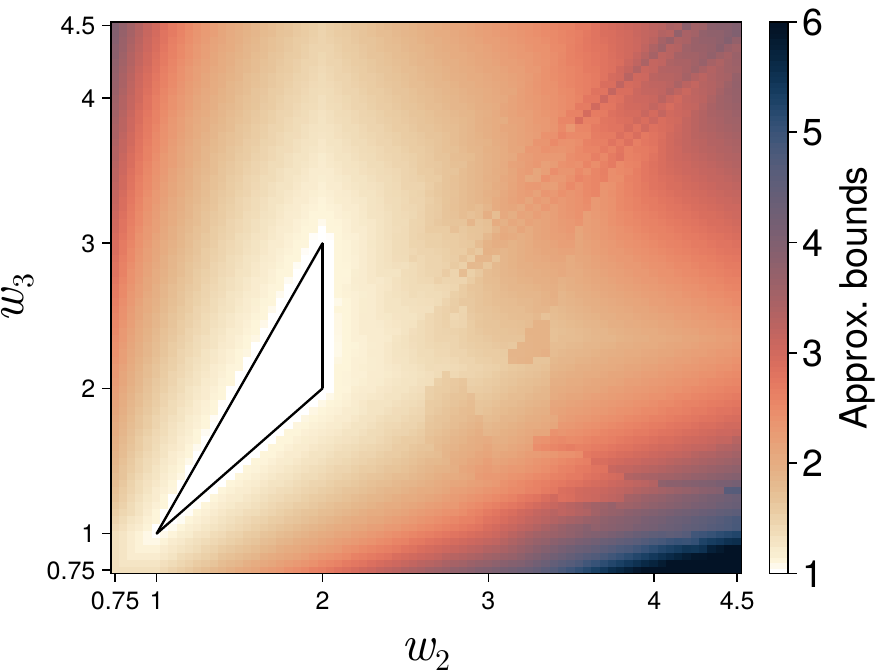} %
		\label{fig:heatmap_Linf}
	}
	\hfill
	\subfigure[Bounds obtained by our new algorithm]{%
		\includegraphics[width=0.45\textwidth]{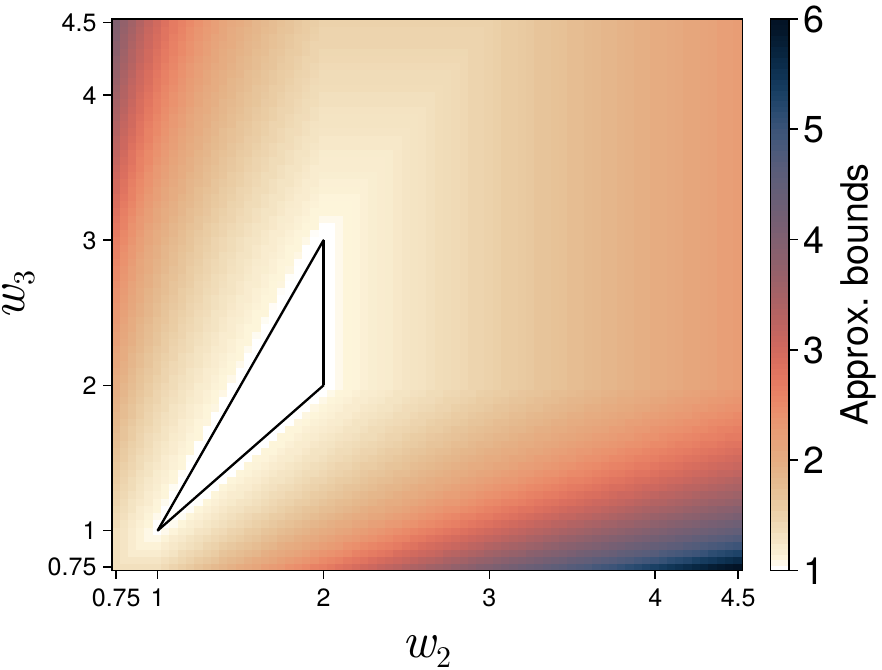} %
		\label{fig:heatmap_plcc}
	}
	\caption{Heatmaps of approximation guarantees obtained for $\cbwa$ when $r \in \{6,7\}$ for a grid of $(w_2,w_3)$ choices when $w_1=1$ is fixed, using four different techniques for projecting on the submodular region. Within the submodular region (shown by a white triangle), the approximation ratio is $1$ as no projection is required. }
	\label{fig:heatmaps_norms}
\end{figure} 
Projecting an arbitrary point onto a convex region is a standard problem in computational geometry. We can use the approximation ratio in Lemma~\ref{lem:projbound} to illustrate how well these standard projection techniques allow us to approximate $\cbwa$. In more detail, we can solve a minimum-norm projection problem of the form
\begin{equation}
	\label{eq:standardproj}
	\min \quad \| \vw - \vw' \| \quad \text{ subject to } \hat{\vw}' \in \mathcal{S}_q
\end{equation}
for some choice of norm $\| \cdot \|.$ The resulting vector may not satisfy $\vw' \geq \vw,$ so we scale it to define a vector $\hat{\vw} = c \vw'$ where $c$ is as small as possible while still satisfying $\hat{\vw} \geq \vw.$ The approximation factor is then given by $\max_i \hat{w}_i/w_i.$
As an illustration, we consider $r \in \{6, 7\}$ and assess the effectiveness of finding a minimum-norm projection of a non-submodular vector onto the submodular space $\mathcal{S}_3$. We specifically consider the $\ell_1$-norm, $\ell_2$-norm, and $\ell_\infty$-norm, as common examples of norms one might wish to minimize. For simplicity, we set $w_1=1$ so that the problem reduces to projecting a two-dimensional point $[w_2, w_3]$ into the submodular region defined by inequalities $w_2 \leq w_3$, $1 \leq w_2 \leq 2$, and $2w_2 \geq w_3 + 1$ (blue region in Figure~\ref{fig:67cb-projection}). It is not hard to show that scaling and projecting in a 2-dimensional space in this way does not affect the resulting approximation guarantee.

\begin{figure}
	\centering
	\subfigure[Difference between $\ell_2$ and $\ell_1$ projections ]{
		\includegraphics[width=0.4\textwidth]{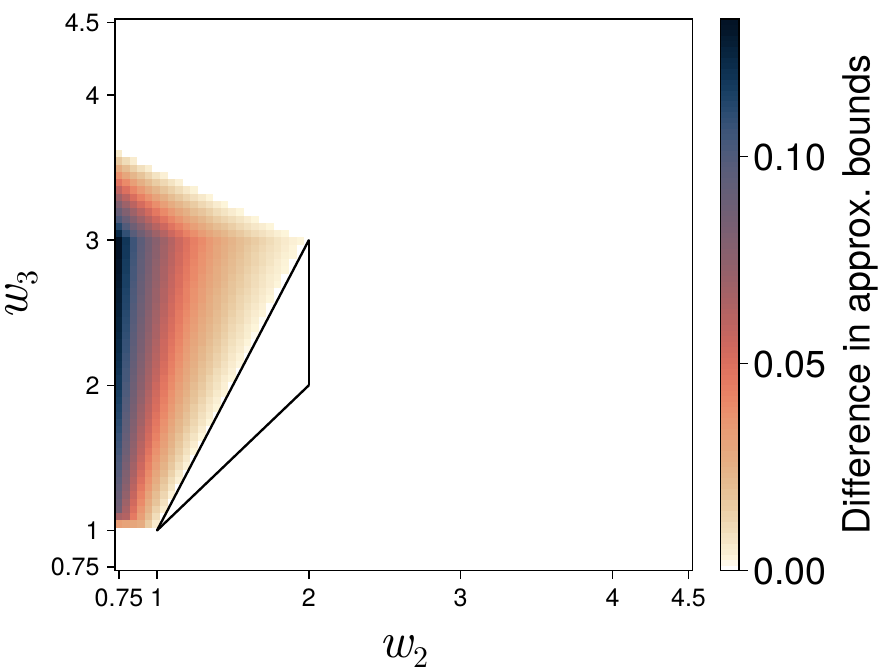}
		\label{fig:heatmap_L1_L2}
	}
	\hfill
	\subfigure[Difference between $\ell_{\infty}$ and $\ell_1$ projections]{
		\includegraphics[width=0.4\textwidth]{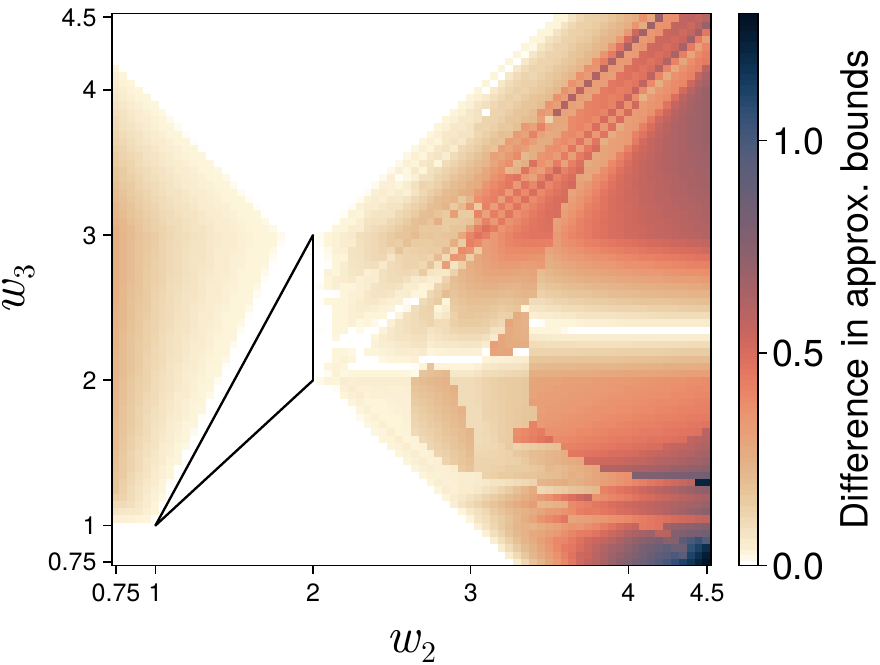} %
		\label{fig:heatmap_L1_Linf}
	}
	\hfill
	\subfigure[Difference between $\ell_{\infty}$ and $\ell_2$ projections]{%
		\includegraphics[width=0.4\textwidth]{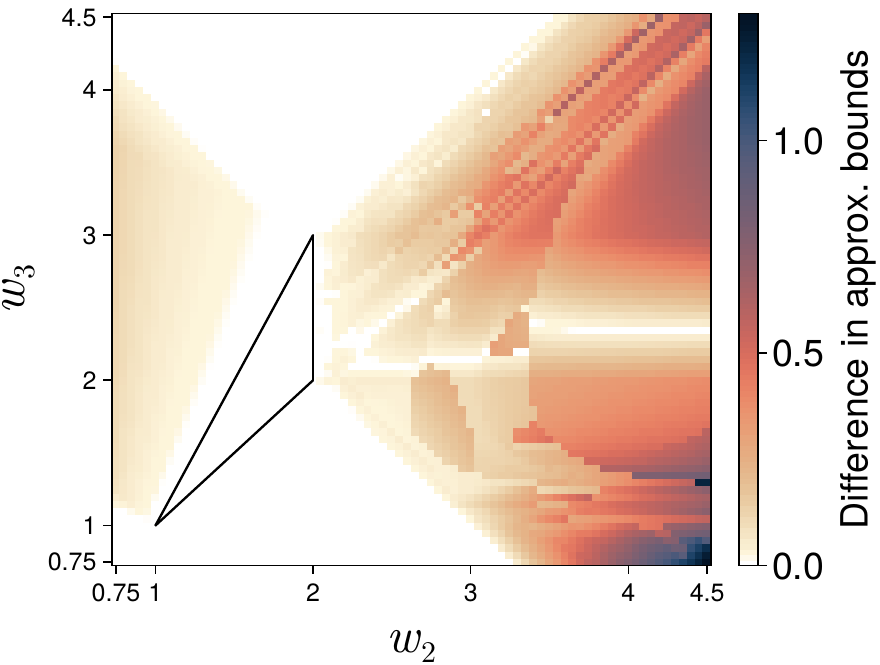} %
		\label{fig:heatmap_Linf_L2}
	}
	\hfill
	\subfigure[Difference between $\ell_1$ and optimal projection]{%
		\includegraphics[width=0.4\textwidth]{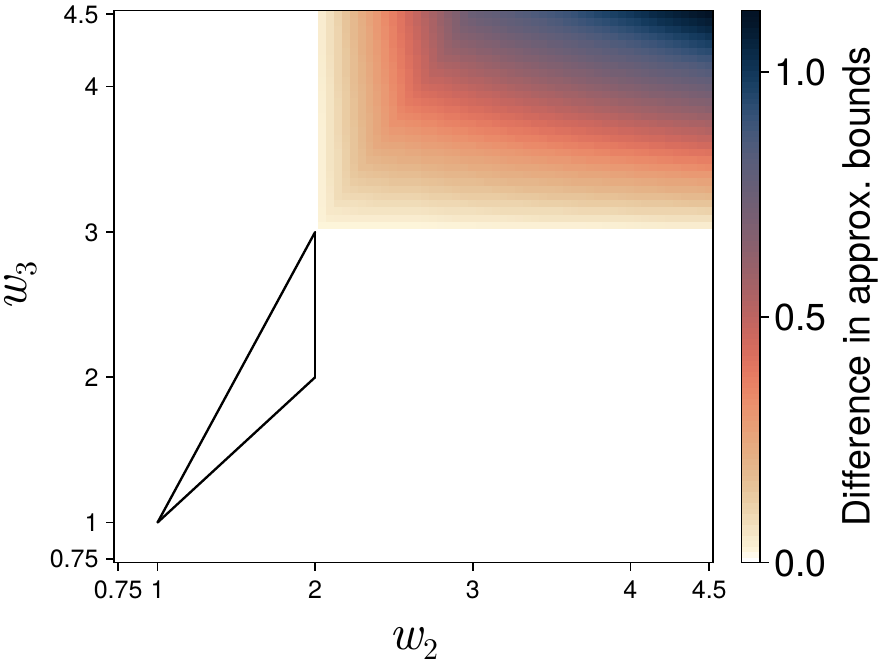} %
		\label{fig:heatmap_L1_Lin}
	}
	\hfill
	\subfigure[Difference between $\ell_2$ and optimal projection]{%
		\includegraphics[width=0.4\textwidth]{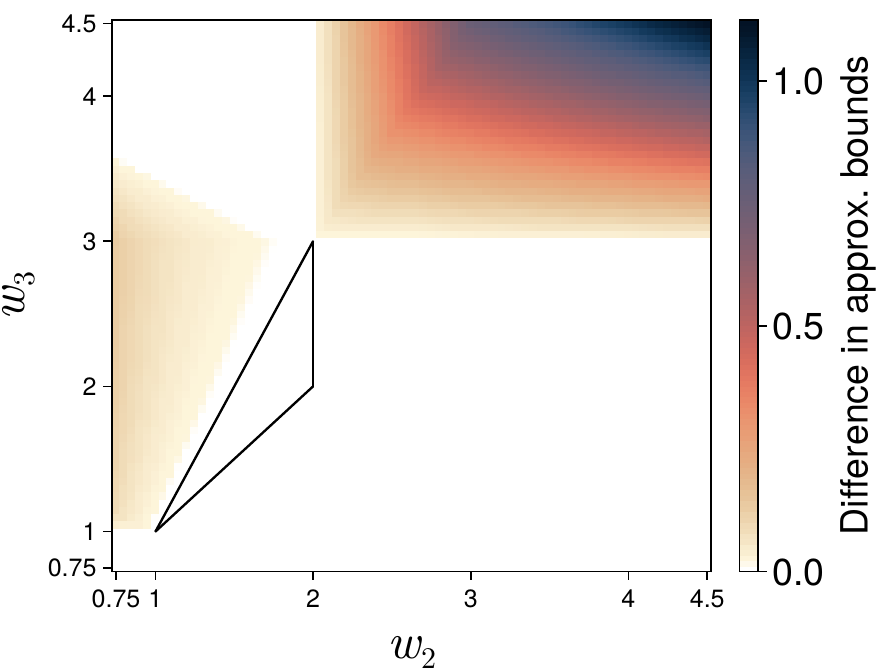} %
		\label{fig:heatmap_L2_Lin}
	}
	\hfill
	\subfigure[Difference between $\ell_{\infty}$ and optimal projection]{%
		\includegraphics[width=0.4\textwidth]{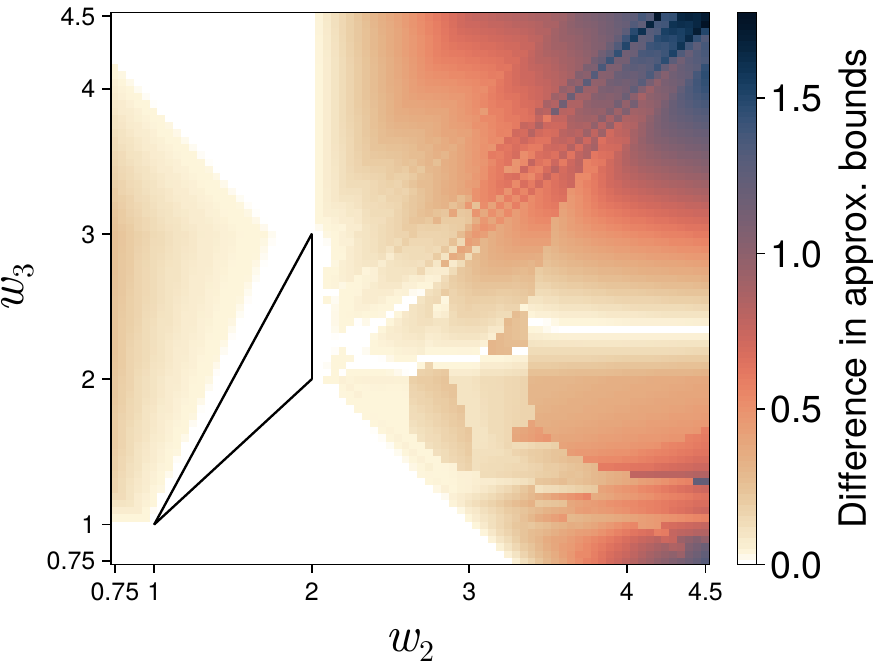} %
		\label{fig:heatmap_Linf_Lin}
	}
	\caption{Heatmaps showing the difference in approximation factors obtained using different types of projection techniques.}
	\label{fig:heatmaps_diffs}
\end{figure}

The best approximation factors for a grid of $(w_2, w_3)$ points is shown using heatmaps in Figures~\ref{fig:heatmap_L1},~\ref{fig:heatmap_L2}, and~\ref{fig:heatmap_Linf}. The darkness of a point indicates how large the worst-case approximation ratio $\rho$ is for a given point $(w_2, w_3)$. Not surprisingly, the approximation factor gets worse as we move farther from the submodular region, for all three norms. The approximation ratios achieved using $\ell_1$ and $\ell_2$ are very similar, though upon close inspection, the $\ell_1$ result is always at least as good as using $\ell_2$, and can be strictly better. The $\ell_\infty$ approximation is never better than $\ell_1$ and $\ell_2$ but can be noticeably worse. Figures~\ref{fig:heatmap_L1_L2} through \ref{fig:heatmap_Linf_L2} display the difference in approximation ratios achieved when using the three different norms. This shows that $\ell_1$ projections produce the best approximation bounds among them, followed by $\ell_2$, and then $\ell_{\infty}$. However, none of these minimum-norm projection approaches produces the best approximation factor. In Figure~\ref{fig:heatmap_plcc} we show the approximation factors achieved by a new projection technique we design, which we will prove provides the optimal approximation factor that can be achieved for $\cbcut(r, \vw)$ by replacing a set of non-submodular splitting penalties with a nearby set of submodular splitting penalties. Figures~\ref{fig:heatmap_L1_Lin}, \ref{fig:heatmap_L2_Lin}, and \ref{fig:heatmap_Linf_Lin} show the difference in approximation factor between each norm-minimizing projection and our new approach. Each norm-minimizing technique produces the same result as our method in \emph{some} regions, but there are also always regions in which they fail to find the best approximation factor. 

\subsection{Finding optimal approximation with convex hull}\label{subsec:convexhull}
We now present full details for our optimal projection method, whose approximation factors are illustrated for $r \in \{6,7\}$ in Figure~\ref{fig:heatmap_plcc}. For simplicity, we will write $w_0 = 0$.
We aim to find new splitting penalties $\hat{w}_i \ge w_i$ that satisfy the submodular constraints while minimizing the largest ratio $\frac{\hat{w}(i)}{w(i)}$ over all $i > 0$. This is the same as solving the following optimization problem,
\begin{align}
	\minimize_{\hat{\vw}} \quad   \kappa &  \quad \label{eq:minmax1}\\
	\text{such that}\quad \kappa & \ge \frac{\hw_i}{w_i}  & i & = 1,2, \hdots, q  \tag{\ref{eq:minmax1}a} \label{eq:constraint_1a}\\
	2\hw_i & \ge \hw_{i-1} + \hw_{i+1}  & i &= 2,3,\hdots,q-1 \tag{\ref{eq:minmax1}b} \label{eq:constraint_1b}\\
	\hw_{i+1} & \ge \hw_i & i & =1,2,\hdots,q-1 \tag{\ref{eq:minmax1}c} \label{eq:constraint_1c}\\  
	\hw_i & \ge w_i & i & = 0,1,2, \hdots, q. \tag{\ref{eq:minmax1}d} \label{eq:constraint_1d}
\end{align}
Since $\vw$ is given, this is just a small linear program (LP). However, we do not need a general LP solver to find the solution. We present a simple approach for finding an optimal $\hat{\vw}$ by casting it as the equivalent task of finding a convex hull of the points $X = \set{(i,w_i) \mid i = 0, \ldots, q}$.

In order to do that let us first define an upper non-decreasing convex hull of points $X$ to be the minimum non-decreasing concave\footnote{We stress that because we are searching for an \emph{upper} convex hull, the resulting function is in fact a \emph{concave} function and not a convex function. It is a known fact that a splitting vector $\vw$ satisfies submodularity constraints in Eq.~\eqref{eq:submodular} if and only if it can be described by $w_i = f(i)$ for some concave function $f$.} function that upper bounds $X$.
Note that since the minimum maintains concavity and monotonicity, the convex hull itself is concave and non-decreasing and upper bounds $X$.
Figure~\ref{fig:plcc_ex} illustrates the convex hull $h$ for a function $\vw$ corresponding to a set of non-submodular splitting penalties.

The following result shows that the convex hull solves \eqref{eq:minmax1}.

\begin{theorem}\label{thm:convexhull}
	Let $h$ be the upper non-decreasing convex hull of $X = \set{(i,w_i) \mid i = 0, \ldots, q}$. Let $\hw_i = h(i)$. Then 
	$\vhw$ solves \eqref{eq:minmax1}.
\end{theorem}

\begin{proof}
	First note that $\vhw$ satisfies the constraints in \eqref{eq:constraint_1b}--\eqref{eq:constraint_1d}. Let $\vw'$ be an optimal solution for \eqref{eq:minmax1}.
	Let $h'$ be the convex non-decreasing function obtained by linearly interpolating the points $\set{(i,w_i') \mid i = 0, \ldots, q}$.
	
	By definition $h = \min(h, h')$.
	Let $\hat{\kappa} = \max \hw_i / w_i$ and $\kappa' = \max w'_i / w_i$. It follows that
	\[
	\hat{\kappa} = \max \frac{h(i)}{w_i} \leq \max \frac{h'(i)}{w_i} = \kappa',
	\]
	proving the claim.
\end{proof}

To find the convex hull we can use 
Andrew's monotone chain algorithm~\citep{andrew1979another}.
The algorithm iteratively constructs the upper convex hull $h$ of $X$ by finding a subset of $X$ that touches $h$. This is done by 
iteratively considering three neighboring points in $X$: if the resulting slopes are not concave, the middle point is deleted. Note that this algorithm does not honor monotonicity. However, we can easily enforce this requirement by adding an additional point $(q + 1, \max w)$ to $X$. Essentially the same algorithm, called Pool Adjacent Violators (PAVA), is also used for solving isotonic regression problems~\citep{leeuw2009isotonic}. The algorithm runs in $O(q)$ time.

\begin{figure}
	\centering
	\resizebox{0.5\textwidth}{!}{ 
		\begin{tikzpicture}
			\begin{scope}[xscale=1.5, yscale=1.0]
				
				
				\draw[->] (0,0) -- (4.5,0) node[right] {$i$};
				\draw[->] (0,0) -- (0,4.0) node[above] {$w$};
				
				\draw[blue, thick] (0,0) -- (2,3);
				\draw[blue, thick] (2,3) -- (3,3);
				\draw[blue, thick] (3,3) -- (4,3);
				
				\draw[red, dashed] (0,0) -- (1,0.7);
				\draw[red, dashed] (1,0.7) -- (2,3);
				\draw[red, dashed] (2,3) -- (3,1.5);
				\draw[red, dashed] (3,1.5) -- (4,2);
				
				\draw[violet, <->] (3,1.6) -- (3,2.9) node[midway, right] {$\rho$};
				
				\node[blue, above] at (1.3,2.2) {$h$};

				\foreach \y/\label in {0/{$0$}, 0.7/{$w_1$}, 1.5/{$w_3$}, 2.0/{$w_4$}, 3.0/{$w_2$}}
				{
					\draw (-0.075,\y) -- (0.075,\y); 
					\node[anchor=east] at (-0.2,\y) {\label}; 
				}
				
				\foreach \x in {0,1,2,3,4}
				\draw (\x,0.1) -- (\x,-0.1) node[below] {\x};
				
			\end{scope}
			
			\filldraw[red] (1.5,0.7) circle (2pt); 
			\filldraw[red] (3,3) circle (2pt);
			\filldraw[red] (4.5,1.5) circle (2pt);
			\filldraw[red] (6,2) circle (2pt);
			\filldraw[green!50!black] (1.5,1.5) circle (2pt); 
			\filldraw[green!50!black] (4.5,3) circle (2pt);
			\filldraw[green!50!black] (6,3) circle (2pt);
			
			\node[green!50!black, above] at (1.5,1.6) {$\hat{w}_1$};
			\node[green!50!black, above] at (3,3) {$\hat{w}_2$};
			\node[green!50!black, above] at (4.5,3) {$\hat{w}_3$};
			\node[green!50!black, above] at (6,3) {$\hat{w}_4$};
			
		\end{tikzpicture}
	}
	\caption{Convex hull $h$ (shown in solid blue) for the integer function $\vw$, with the discrete values of $\vw(i)$ marked in red. The blue line segments for each interval $[i,i+1]$ represent the linear pieces of $h$, while the largest gap between $h(i)$ and $\vw(i)$ (here at $i=3$) is denoted by the approximation bound $\rho$. The red dashed line interpolates $\vw$; the fact that it is not a concave function indicates that $\vw$ does not define splitting penalties for a submodular function.}
	\label{fig:plcc_ex}
\end{figure}

\section{Approximation Hardness Results}
\label{sec:approxhard}
We now prove hardness results to strongly indicate that our projection-based technique for non-submodular \cbcut{} produces the best approximation guarantees we can hope for. We not only prove APX-hardness for the non-submodular case but also show a large set of cases (including all cases for non-submodular \cbcut{} in 4-uniform hypergraphs) where our projection-based approximation techniques in Section~\ref{sec:projections} are the best possible assuming the Unique Games Conjecture.

\subsection{APX-hardness via reduction from \textsc{MaxCut}}
Let $G = (V,E)$ represent an instance of \textsc{MaxCut} where $m = |E|$ and where we use $k^*$ to denote the optimal number of edges that are cut. For our APX-hardness results, we will use the fact that there is always a way to cut at least half the edges in a graph, so $k^* \geq m/2$. We also use the fact that \textsc{MaxCut} is NP-hard to approximate to within a factor better than $16/17$~\citep{hastad}.

\begin{lemma}
	Assume a splitting vector $\vw$ with
	$2w_i < w_{i - 1} + w_{i + 1}$, for some $i$.
	Then it is NP-hard to approximate $\cbwa$ to within a factor smaller than 
	\begin{equation*}
		1+ \frac{w_{i - 1} + w_{i + 1} - 2w_i}{17(w_{i - 1} + w_{i + 1}) + 34w_i}.
	\end{equation*}
\end{lemma}
\begin{proof}
	Consider the reduction from \textsc{MaxCut} to $\cbwa$ given in the proof of Theorem~\ref{thm:nphard}. 
	
	Let us write $u = w_{i - 1} + w_{i + 1}$.
	A bipartition of the nodes in $V$ that cuts $k$ edges corresponds to an $s$-$t$ cut in the reduced hypergraph with cut value
	$2kw_i + (m-k)u$.
	
	Assume that we have a polynomial-time $\beta$-approximation algorithm for $\cbwa$. If the sets $S$ or $T$ are cut, then according to the proof, the resulting cost is at least $\alpha$, implying that $\beta \geq 2$, in which case we have nothing to prove. Consequently, an $s$-$t$ cut corresponds to a cut in $G$ with $k'$ edges.
	
	The approximation guarantee implies
	\[
	2k'w_i + (m-k')u \leq \beta (2k^*w_i +(m-k^*)u).
	\]
	Rearranging the terms leads to
	\[
	\beta(u-2w_i)k^* \leq (\beta - 1)mu + (u-2w_i)k'\leq (\beta - 1)2k^*u + (u - 2w_i)k',
	\]
	where in the last step we have used the fact that $k^* \ge m/2$.
	Rearranging the terms and applying the fact that $k'/k^* \leq 16/17$ leads to
	\[
	\beta(u - 2w_i) - 2(\beta -1)u \le {\frac{k'}{k^*}}(u-2w_i) \leq \frac{16}{17}(u - 2w_i).
	\]
	Solving for $\beta$ yields
	\[
	\beta \ge \frac{18u + 32w_i}{17(2w_i+u)}= \frac{18u + 32w_i}{17u + 34w_i} = 1+ \frac{w_{i - 1} + w_{i + 1} - 2w_i}{17(w_{i - 1} + w_{i + 1}) + 34w_i},
	\]
	proving the claim.
\end{proof}
We prove a similar result when splitting weights are non-increasing.
\begin{lemma}
	Assume a weight vector $\vw$ with
	$w_i > w_{i + 1}$, for some $i$.
	Then it is NP-hard to approximate $\cbwa$ to within a factor smaller than 
	\begin{equation*}
		1 + \frac{w_i -w_{i + 1}}{17(w_i+w_{i + 1})}.
	\end{equation*}
\end{lemma}
\begin{proof}
	Consider the reduction from \textsc{MaxCut} to $\cbwa$ given in the proof of Theorem~\ref{thm:nphard}. 
	
	Cutting $k$ edges in the \textsc{MaxCut} instance $G = (V,E)$ corresponds to a hypergraph $s$-$t$ cut value of $kw_{i + 1} + (m-k)w_i$ (recall that $m = |E|$).
	
	Assume that we have a polynomial-time $\beta$-approximation algorithm for $\cbwa$. If the sets $S$, $T$, or $U_x$ are cut, then according to the proof, the resulting cost is at least $\alpha$, implying that $\beta \geq 2$, in which case we have nothing to prove. Consequently, an $s$-$t$ cut corresponds to a cut in $G$ with $k'$ edges.
	
	The approximation guarantee implies
	\[
	k'w_{i + 1} + (m - k')w_i \le \beta(k^*w_{i + 1} + (m - k^*)w_i).
	\]
	Rearranging the terms and using the fact that $k^* \geq m/2$ leads to
	\[
	\beta k^*(w_i - w_{i + 1}) \le k'(w_i -w_{i + 1}) + (\beta - 1)m w_i \le k'(w_i -w_{i + 1}) + 2(\beta - 1)k^*w_i.
	\]
	Further rearranging and using the fact that $k'/k^* \leq \frac{16}{17}$ leads to
	\[
	\beta(w_i- w_{i + 1}) - 2(\beta -1)w_i \le \frac{k'}{k^*}(w_i-w_{i + 1}) \le \frac{16}{17}(w_i-w_{i + 1}).
	\]
	Solving for $\beta$ leads to
	\[
	\beta \ge 1 + \frac{w_i -w_{i + 1}}{17(w_i+w_{i + 1})},
	\]
	proving the claim.
\end{proof}
For $4$-uniform hypergraphs, these lemmas tell us that for any fixed $w_2 \notin [1,2]$, there exists an $\epsilon > 0$ (depending on $w_2$) such that finding an approximation within a factor $1+\epsilon$ is NP-hard. As $w_2$ gets further from the submodular region (i.e., taking a limit $w_2 \rightarrow 0$ or $w_2 \rightarrow \infty$), the approximation factor gets worse. However, even for the most extreme values of $w_2$, this only rules out the possibility of obtaining an approximation better than ${18}/{17}$. In contrast, our best approximation factors become arbitrarily bad as $w_2$ goes to zero or infinity. We would like, therefore, to tighten this gap to show that the best approximation factors also get arbitrarily bad as we get further from submodularity.

\subsection{Asymptotic inapproximability for large $w_i$}
In prior sections, we treated splitting weights as fixed constants. We now briefly consider a more general setting where the splitting weights $w_i$ are allowed to grow in terms of the size of the number of nodes or edges in the hypergraph. We show that for extreme values of $w_i$, our approximation guarantees are asymptotically tight. Our results in this section also specifically show hardness and inapproximability results for the \textsc{No-Even-Split} problem.

\begin{theorem}
	\label{thr:asymptapprox}
	Assume that there is an algorithm for $\cbcut(r, \vw)$ with $w_1 = 1$, yielding an approximation guarantee of $c(n, \vw)$, where $n$ is the number of nodes in the hypergraph. Assume that the algorithm runs in polynomial time in $n$.
	Select $i \geq 2$.
	Then, barring $\mathit{P} = \mathit{NP}$, we have $c(n, \vw) \notin O(\min(n, w_i)^{1 - \epsilon})$, for any $1 > \epsilon > 0$. 
\end{theorem}

We will reduce 3SAT to $\cbcut(r, \vw)$ to prove the result. An instance $\phi$ of 3SAT consists of a conjunction (ANDs) of clauses, where each clause is a disjunction (ORs) of exactly three literals. A literal is either a boolean variable or its negation. The question is if there exists an assignment of the variables as true or false such that the conjunction $\phi$ is logically true.

Given a 3SAT instance $\phi$ with $M$ clauses and $N$ variables, we construct an $r$-uniform hypergraph $\mathcal{H}$ in the following way. Let $\alpha$ denote a positive integer that we will define later. Add two sets of nodes $S = \{s_1,s_2,\ldots,s_{\alpha}\}$ and $T = \{t_1,t_2,\ldots,t_{\alpha}\}$, consisting of all possible hyperedges of size $r$. These edges aim to ensure that all nodes in $S$ are on the same side of the cut, by setting $\alpha$ large enough. Similarly, all nodes in $T$ should be on the same side. Let $s=s_1$ be the source and $t=t_1$ the sink of our cut problem.

For each variable $x$ create two sets of $\alpha$ nodes $\{x_1,x_2,\ldots,x_{\alpha}\}$ and $\{\neg x_1,\neg x_2,\ldots,\neg x_{\alpha}\}$, again with all possible hyperedges of size $r$ within the sets. In addition, create hyperedges 
\[
(s_1,\ldots,s_{r-i}, x_1, \neg x_1, \ldots, \neg x_{i-1}) \quad\text{and}\quad (t_1,\ldots,t_{r-i}, x_1, \neg x_1, \ldots, \neg x_{i-1}).
\] 
Then having no $(i, r-i)$-splits ensures $x_1$ and $\neg x_1$ are not on the same side of the cut, and having $x_1$ and $\neg x_1$ not on the same side means we do not have a $(i, r-i)$-split.

For each clause $j$, create another set of $\alpha$ nodes $\{z_j^1, \ldots, z_j^\alpha\}$ with all hyperedges of size $r$ within the set to ensure they are on the same side of the cut, and denote the corresponding truth value as $z_j$.
Then, assuming clause $j$ is $a \lor b \lor c$, create the hyperedges
\begin{equation} \label{eq:hyperedges1}
	(s_1, \neg a_1, \ldots, \neg a_{i-1}, \neg b_1, z_j^1, \ldots, z_j^{r-i-1})
\end{equation}
and
\begin{equation} \label{eq:hyperedges2}
	(s_1, c_1, \ldots, c_{i-1}, z_j^1, \ldots, z_j^{r-i}).
\end{equation}

We have the following lemma.

\begin{lemma}
	\label{lem:nphardnesslargew_i}
	Assume $\alpha \geq \max(10, 2r)$.
	If $\phi$ is satisfiable, then there is an $s$-$t$ cut with cost at most $(w_{i-1}+1)(N+2M)$.
	If $\phi$ is not satisfiable, any cut will cost at least $\min(\alpha, w_i)$.
\end{lemma}
\begin{proof}
	As in Section~\ref{sec:maxcut_reductions}, when $\alpha \geq \max(10, 2r)$, cutting any of the sets with $\alpha$ nodes would yield a cost of at least $\alpha$.

	Suppose we have a cut of cost smaller than $\min(\alpha, w_i)$.
	This implies that there are no $(i, r-i)$ cuts and none of the sets of $\alpha$ nodes are cut. 
	
	We construct a satisfying assignment by setting the literals on the side of $t_1$ as true and the literals on the $s_1$ side as false. Consider a clause $j$. For edges in Eq.~\eqref{eq:hyperedges1}, having no $(i, r-i)$-split ensures that if $\neg a$ and $\neg b$ are true, so is the clause $j$.
	For edges in Eq.~\eqref{eq:hyperedges2}, having no $(i, r-i)$-split ensures that if $c$ is false, so is the clause $j$.
	In summary, if we have no $(i, r-i)$-splits, it is impossible for $a$, $b$, and $c$ to all be false, as then $j$ would be true by Eq.~\eqref{eq:hyperedges1} but false by Eq.~\eqref{eq:hyperedges2}, which is a contradiction. Therefore, the clause $j$ is satisfied.
	This holds for each clause $j$, so $\phi$ is satisfiable.
	
	Conversely, if $\phi$ is satisfiable, we may split the literals according to a satisfying assignment with true literals on the $t_1$ side and false literals on the $s_1$ side. 
	Assign $z_j = \neg a \land \neg b$. Then note that the hyperedge in Eq.~\eqref{eq:hyperedges1} will not be a $(i, r-i)$-split because if $z_j$ and $\neg a$ are on the same side, the minority side will have at most one node because it is not possible for $\neg b$ to be with $s_1$ while $z_j$ is true. Furthermore, the only case where $z_j$ and $\neg a$ are on different sides is when $\neg a$ is true and $\neg b$ is false, which implies we have a $(i-1, r-i+1)$-split.
	Also note that the hyperedge in Eq.~\eqref{eq:hyperedges2} will not be a $(i, r-i)$-split because the only way to have an $(i, r-i)$-split would be for $c$ to be false while $z_j$ is true, but this would mean $a$, $b$, and $c$ are all false, which is a contradiction since the assignment must satisfy every clause.
	
	We can also see that when $\phi$ is satisfiable, we have a solution where all the splits will be $(1, r-1)$-splits or $(i-1, r-i+1)$-splits. We have $2N$ hyperedges for the $N$ variables, where one hyperedge will yield a cost of $w_1$ while the other will yield a cost of $w_{i-1}$. In addition, we have $2M$ hyperedges for the $M$ clauses, where each yields a cost of at most $\max(w_1, w_{i-1}) < w_{i-1}+1$. 
	We then have a total cost of at most 
	\[
	N(w_{i-1}+1) + 2M\max(w_1, w_{i-1}) < (w_{i-1}+1)(N+2M),
	\]
	proving the claim.
\end{proof}

We now argue that the above reduction also induces a gap in the optimum value. We will show that we cannot have an approximation factor that is asymptotically better than $w_i$ and $n$ for sufficiently large $w_i$ and $n$.

\begin{proof}[Proof of Theorem~\ref{thr:asymptapprox}]
	
	Fix $0 < \epsilon < 1$. Assume that we have an approximation algorithm with an approximation guarantee of $c(n, \vw) \in O(\min(w_i,n)^{1 - \epsilon})$. Then there are constants $\ell$ and $\gamma$ such that when $w_i \geq \ell$ and $n \geq \ell$ we have $c(n, \vw) < \gamma \min(w_i,n)^{1 - \epsilon}$.
	
	Select $w_2, \ldots, w_{i - 1}$;
	here any choice works as long as $w_{i-1}$ is of polynomial size of $N$, for example $w_2 = \cdots = w_{i - 1} = 1$.
	Let $k = (w_{i-1}+1)(N+2M)$ and $b = 2N+M+2$. Set $w_i = \max((\gamma k)^{1/\epsilon}, \ell)$
	and $\alpha = \ceil*{\max((b^{1-\epsilon}\gamma k)^{1/\epsilon}, 2r, 10, \ell)}$. Note that the hypergraph contains $n = \alpha b$ nodes, and $\gamma k \leq w_i^\epsilon$ and
	$\gamma k b^{1 - \epsilon} \leq \alpha^\epsilon$. Select also $w_{i + 1}, \ldots, w_q$. Here any choice works, for example $w_{i + 1} = \cdots = w_q  = w_i$.
	
	Let $A$ be the cut cost returned by the approximation algorithm, and let $\mathit{OPT}$ be the optimal cut cost.
	
	We claim that $\phi$ is satisfiable if and only if $A < \min(\alpha, w_i)$. If $\phi$ is not satisfiable, then Lemma~\ref{lem:nphardnesslargew_i} implies the claim. If $\phi$ is satisfiable, then Lemma~\ref{lem:nphardnesslargew_i} implies
	\begin{align*}
		A & \leq c(n, \vw)\mathit{OPT} \leq c(n, \vw)k < \gamma k w_i^{1 - \epsilon} \leq w_i \quad\text{and} \\
		A & \leq c(n, \vw)\mathit{OPT} \leq c(n, \vw)  k < \gamma k n^{1 - \epsilon}
		= \gamma k b^{1-\epsilon} \alpha^{1 - \epsilon} \leq  \alpha.
	\end{align*}
	Therefore, $\phi$ is satisfiable if and only if $A < \min(\alpha, w_i)$, which means that the approximation algorithm can be used to solve 3SAT in polynomial time. Assuming $\mathit{P} \neq \mathit{NP}$, this is a contradiction, so any approximation factor must be $c(n, \vw) \notin O(\min(n, w_i)^{1 - \epsilon})$.
\end{proof}

The result holds even if we require $\vw$ to be contained in some specific set of splitting functions $\vW$. For example, it holds if $\vW$ is the set of splitting functions where $w_i$ is arbitrarily large but $w_1, \ldots, w_{i-1}, w_{i+1}, \ldots w_q$ are fixed constants. It also holds if we set $\vW$ to be all monotonic weight vectors where $w_1, \ldots, w_{i-1}$ are fixed constants but the remaining splitting weights are arbitrarily large. To prove the claim, we need to modify the proof by selecting $\vw \in \vW$.

We can further extend Theorem~\ref{thr:asymptapprox} by allowing the running time of the algorithm to depend on $\vw$. The claim follows if we choose $\vw$ to be of polynomial size of $N$ in the proof.

The proofs of
Lemma~\ref{lem:nphardnesslargew_i} and Theorem~\ref{thr:asymptapprox} also imply the following result about \textsc{No-Even-Split}, where $r=4$ and even splits are forbidden by setting $w_2 = \infty$.
\begin{corollary}
	\textsc{No-Even-Split} is $\text{NP}$-hard and does not have any approximation algorithm with an approximation guarantee that is asymptotically better than $O(n^{1 - \epsilon})$, for any $\epsilon > 0$, unless $\text{P}=\text{NP}$.
\end{corollary}
This answers the question about the complexity of \textsc{No-Even-Split} raised in the list of open questions in applied combinatorics of~\cite{aksoy2023seven}.

\subsection{Tighter inapproximability via unique games conjecture}
To prove stronger approximation results, we again leverage the connection to Valued Constraint Satisfaction Problems. Every VCSP permits a linear programming (LP) relaxation known as the Basic LP~\citep{thapper2012power}, which we will define shortly for VCSPs corresponding to hypergraph $s$-$t$ cut problems.

\citet{ene2013local} showed that for certain constraint languages, the integrality gap of the Basic LP imposes UGC-hardness on achieving approximation ratios better than that gap. The result however relies on the presence of constraints that correspond to the cut function on standard edges (i.e., size-2 hyperedges). To leverage these results and prove the UGC-hardness bounds for approximating $\wcbcut(r,\vw)$, we first establish the approximation equivalence between $\wcbcut(r, \vw)$ and a variant of the problem that includes standard edges in addition to size-$r$ hyperedges.

Formally we are given a hypergraph $\mathcal{H} = (\V, \E \cup E)$ where $\E$ is a set of (scalar-weighted) hyperedges of size-$r$, and $E$ is a set of weighted edges. The $\ecbcut(r, \vw)$  problem is given by
\begin{align}
	\label{eq:cardwithedges}
	\minimize \quad 
	\cut_E(S) + \sum_{i=1}^{q} w_i \cdot C(\partial S_i) \quad \text{ subject to } s \in S \text{ and } t \in \bar{S},
\end{align}
where $\partial S_i$ denotes the size-$r$ hyperedges that are $(i, r-i)$ split, and $\textbf{cut}_E(S)$ is the standard graph cut function for the graph defined by edges $E$. 
\begin{lemma}
	There is a polynomial-time $c(\vw)$-approximation for $\ecbcut(r, \vw)$ if and only if there is a polynomial-time $c(\vw)$-approximation algorithm for $\wcbcut(r, \vw)$.
\end{lemma}
\begin{proof}
	$\wcbcut(r, \vw)$ is a special case of $\ecbcut(r, \vw)$ where $E = \emptyset$. We can reduce the latter to the former in an approximation-preserving way by replacing each $(x,y) \in E$ with a size-$r$ hyperedge $e_{xy} = (x,y, a_1, \hdots, a_{r-2})$, where $\{a_1, \hdots, a_{r-2}\}$ are new nodes that only show up in this hyperedge. In the resulting instance of $\wcbcut(r, \vw)$, if $x$ and $y$ are placed on the same side of a cut, the minimum cut penalty for $e_{xy}$ is obtained by placing the nodes $\{a_1, \hdots, a_{r-2}\}$ on that same side. This results in a cut penalty of 0, as would be the case for the standard edge cut penalty for $(x,y)$. If $x$ and $y$ are separated, then $\{a_1, \hdots, a_{r-2}\}$ will arrange themselves in a way that leads to a penalty of $\min_{i \in [q]} w_i > 0$. We can scale this penalty by a scalar weight to ensure the resulting penalty at $e_{xy}$ exactly coincides with the weight of $(x,y)$ in the instance of $\ecbcut(r, \vw)$.
\end{proof}
Given this result, we know that any approximation hardness results we prove for $\ecbcut(r, \vw)$ will also hold for $\wcbcut(r, \vw)$.

\paragraph{The Basic LP for hypergraph $s$-$t$ cut problems}
Having established the relationship between VCSPs and generalized hypergraph $s$-$t$ cut problems in Sections~\ref{sec:prelims} and~\ref{sec:nphard}, we will go back and forth between the two views interchangeably. For a generalized hypergraph $s$-$t$ cut problem on $\mathcal{H} = (\V, \E \cup E)$ (whether or not splitting functions are cardinality-based), the Basic LP is given by
\begin{align}
	\label{eq:basiclp}
	\text{minimize} \quad & \sum_{e \in \E \cup E} c_e \cdot \sum_{A \subseteq e} y_{e,A} \cdot \vw_e(A) \quad & \\
	\label{lp1}
	\text{subject to} \quad & x_{v,s} = \sum_{A \subseteq e : v \in A} y_{e,A},  &  e \in \E \cup E ,  v \in e, \\
	\label{lp2}
	&x_{v,t} = \sum_{A \subseteq e : v \in e \setminus A} y_{e,A}, &  e \in \E\cup E ,  v \in e,  \\
	&0\le y_{e,A}\le 1 ,&  e\in \E\cup E, A \in e,\\
	& x_{v,s} + x_{v,t} = 1, &   v \in \V,  \\
	& 0 \le x_{v,s} \le 1 \text{ and } 0 \le x_{v,t} \le 1, &  v \in \V, \\
	& x_{s,s} = 1 \text{ and } x_{t,t} = 1.
\end{align} 
Here, $c_e \geq 0$ is a scalar weight associated with each $e \in \E \cup E$.  We use variables $x_{v,s}$ and $x_{v,t}$ for each node $v \in \V$ to indicate the fractional assignment of $v$ to the $s$ and $t$ sides. If we restricted variables to be binary, then the solution would exactly be the optimal solution for the generalized hypergraph $s$-$t$ cut problem. Hence, the solution to the LP lower bounds the optimal $s$-$t$ cut value.

This Basic LP for hypergraph $s$-$t$ cut problems can be easily derived from the more general presentation of the Basic LP in Section 4 of~\citet{ene2015local} (the online full version of an earlier conference paper~\citep{ene2013local}).
The more general LP includes an LP variable $x_{v,\ell} \in [0,1]$ for every VCSP variable $v$ and possible assignment $\ell$ for $v$. For Boolean VCSPs, $\ell \in \{0,1\}$, though the Basic LP is also defined for non-Boolean VCSPs. The other variables and constraints in the LP are formulated in such a way that the optimal solution to the LP lower bounds the optimal solution for the VSCP instance. The optimal value for $x_{v,\ell}$ can be interpreted as the probability of assigning variable $v$ to label $\ell$. 

The hardness result by~\citet{ene2015local} for VCSPs translates to the following result for the above Basic LP for generalized hypergraph $s$-$t$ cut problems. 

\begin{theorem}
	\label{thm:ene} 
	If $\mathcal{H}$ is an instance of $\ecbcut{}(r,\vw)$ and $c < \mathit{OPT}(\mathcal{H})/\mathit{LP}(\mathcal{H})$, it is UGC-hard to approximate $\ecbcut{}(r,\vw)$ to within a factor $c$, where $\mathit{OPT}(\mathcal{H})$ denotes the optimal solution for this instance and $\mathit{LP}(\mathcal{H})$ denotes the value of the Basic LP.
\end{theorem}
\begin{proof}
	This result can be viewed as a simple corollary of Theorem 4.3 by \citet{ene2015local}, which states that when the constraint language includes the two-variable \emph{not-all-equal} predicate, then under the Unique Games Conjecture the integrality gap of the Basic LP relaxation lower bounds the best possible approximation factor for the corresponding valued constraint language. This predicate on two variables, denoted by $\text{NAE}_2$ is defined as  
	\begin{align*}
		\text{NAE}_2(x,y) =
		\begin{cases} 
			0 & \text{if } x = y \\
			1 & \text{if } x \neq y.
		\end{cases}
	\end{align*}
	When we translate the $\ecbcut(r,\vw)$ problem into the VCSP framework using the reduction described in Section~\ref{sec:vcsp}, we see that the $\ecbcut(r, \vw)$ problem is equivalent to the constraint language with cost functions $\{\phi_r, \phi_s, \phi_t, \phi_{st}, \text{NAE}_2\}$.
	
	Here the standard graph cut function for an edge corresponds to the $\text{NAE}_2$ predicate. This connection allows us to directly apply the hardness results of Theorem 4.3 by \citet{ene2015local} to the $\ecbcut(r,\vw)$ problem.
	Therefore, given an instance $\mathcal{H}$ with an integrality gap of $OPT(\mathcal{H})/LP(\mathcal{H})$, it is UGC-hard to approximate $\ecbcut(r,\vw)$ within a factor $c< OPT(\mathcal{H})/LP(\mathcal{H})$.
\end{proof}

Theorem~\ref{thm:ene} states that in order to prove UGC-hardness results, we just need to show a relevant integrality gap example for the Basic LP.

\paragraph{Integrality gap for $w_2 < 1$}

\begin{figure}  
	\subfigure[Optimum solution with $OPT =1$]  
	{  
		\label{fig:igap_opt_case1}
		\begin{tikzpicture}[scale=.45]  
			

			\definecolor{lightorange}{RGB}{255,204,153}
			\definecolor{lightyellow}{RGB}{255,255,153}
			\definecolor{gray}{RGB}{200,200,200}
			\definecolor{maroon}{rgb}{0.5, 0, 0} 
			
			\node[circle, fill=lightorange, draw, minimum size=0.5cm] (s) at (1,4.5) {$s$};
			\node[circle, fill=gray, draw, minimum size=0.5cm] (1) at (3.5,6.0) {$1$};
			\node[circle, fill=gray, draw, minimum size=0.5cm] (2) at (3.5,3) {$2$};
			\node[circle, fill=lightorange, draw, minimum size=0.5cm] (t) at (7,4.5) {$t$};
			
			\draw[thick,maroon] (1) -- (2);
			
			\draw[rounded corners=9pt,thick,maroon] (0,2) rectangle (8,7);  
			\draw [thick, dashed] (2,1) to[line to] (2,8); 
			
			\draw[maroon] (-0.5,2.5) node {$g$}; 
			\draw[maroon] (-1.0,1.5) node {$c_g=1$};
			\draw[maroon] (4.0,4.5) node {$2$};

			\draw (0.2,7.5) node {$x_{s,s}=1$};
			
			\draw (5.5,8.0) node {$x_{1,s}=0$};
			\draw (4.0,1.0) node {$x_{2,s}=0$};
			
			\draw (8.0,1.0) node {$x_{t,s}=0$};
		\end{tikzpicture}  
	} \hfill
	\subfigure[Basic LP solution with value $w_2$] { 
		\label{fig:igap_blp_case1}
		\begin{tikzpicture}[scale = 0.45]
			

			\definecolor{lightorange}{RGB}{255,204,153}
			\definecolor{lightyellow}{RGB}{255,255,153}
			\definecolor{gray}{RGB}{200,200,200}
			\definecolor{maroon}{rgb}{0.5, 0, 0} 
			
			\node[circle, fill=lightorange, draw, minimum size=0.5cm] (s) at (1,4.5) {$s$};
			\node[circle, fill=gray, draw, minimum size=0.5cm] (1) at (3.5,6.0) {$1$};
			\node[circle, fill=gray, draw, minimum size=0.5cm] (2) at (3.5,3) {$2$};
			
			\node[circle, fill=lightorange, draw, minimum size=0.5cm] (t) at (7,4.5) {$t$};
			
			\draw[thick,maroon] (1) -- (2);
			
			\draw[rounded corners=9pt,thick,maroon] (0,2) rectangle (8,7);

			\draw[maroon] (-0.5,2.5) node {$g$}; 
			\draw[maroon] (-1.0,1.5) node {$c_g=1$};
			\draw[maroon] (4.0,4.5) node {$2$};

			\draw (0.4,7.5) node {$x_{s,s}=1$};
			\draw (5.5,8.0) node {$x_{1,s}=0.5$};
			\draw (3.0,1.0) node {$x_{2,s}=0.5$};
			\draw (8.0,1.0) node {$x_{t,s}=0$};
	\end{tikzpicture}  }
	\caption{Integrality gap instance of $\ecbcut(4,(w_1=1,w_2<1))$}
	\label{fig:igap_case1}
\end{figure}

Consider an instance $\mathcal{H} = (\V ,\E \cup E)$ of $\ecbcut{}(4,(w_1 = 1, w_2 < 1))$ with four nodes $\V = \{1,2,s,t\}$, an edge $(1,2)$ and a hyperedge $g = (s,t,1,2)$ as shown in Figure~\ref{fig:igap_case1}. The hyperedge $g$ has a weight $c_g = 1$, while the edge $(1,2)$ has a weight $2w_1 = 2$. A minimum $s$-$t$ cut has a value of $\mathit{OPT}(\mathcal{H}) = 1$, which can be obtained by placing node $s$ on its own or $t$ on its own. 
Figure \ref{fig:igap_opt_case1} illustrates an optimal solution where node $s$ is placed by itself, along with the binary feasible variables for the Basic LP that represents this $s$-$t$ cut.

The Basic LP has a feasible fractional solution for the given instance where for each non-terminal node $v$ we have $x_{v,s} = x_{v,t} = 0.5$ (see Figure~\ref{fig:igap_blp_case1}). 
For variables $y_{(1,2),A}$ where $A \subseteq (1,2)$, we can define $y_{(1,2),\emptyset} = y_{(1,2),\{1,2\}} = 0.5 $ and set $y_{(1,2),A} = 0$ for every $A \notin \{\emptyset, \{1,2\} \}$.  For node $v \in (1,2)$ we have
\begin{align*}
	x_{v,s} &= 0.5 = y_{(1,2),\{1,2\}} = \sum_{A \subseteq (1,2) \colon v \in A} y_{(1,2),A}, \\
	x_{v,t} &= 0.5 = y_{(1,2), \emptyset} = \sum_{A \subseteq (1,2) \colon v \in (1,2)\backslash A} y_{(1,2),A},
\end{align*}
so we see that constraints in~\eqref{lp1} and~\eqref{lp2} are satisfied for edge $(1,2)$. 

For $ g = (s,t,1,2)$, set $y_{g,\{s,1\}} = y_{g,\{s,2\}} = 0.5$ and $y_{g,A} = 0$ for every $A \notin \{\{s,1\}, \{s,2\} \}$. We can confirm that the constraints in~\eqref{lp1} and~\eqref{lp2} are also satisfied for $g$,
\begin{align}
	\begin{array}{ll}
		x_{1,s} = y_{g,\{s,1\}} + 0 = 0.5  &\quad x_{1,t} = y_{g,\{s,2\}} + 0 = 0.5\\
		x_{2,s} = y_{g,\{s,2\}} + 0 = 0.5  &\quad x_{2,t} = y_{g,\{s,1\}} + 0 = 0.5\\
		x_{s,s} = y_{g,\{s,1\}} + y_{g,\{s,2\}} = 1.0  &\quad x_{s,t} = 0 \\
		x_{t,t} = y_{g,\{s,1\}} + y_{g,\{s,2\}} = 1.0   &\quad x_{t,s} = 0.
	\end{array}
\end{align}
Observe that $y_{g,A}\cdot\vw_g(A) = 0$ for every $A \notin \{\{s,1\},\{s,2\}\}$. The LP value for this feasible solution (which is in fact optimal for the LP) is therefore
\begin{align*}
	\mathit{LP}(\mathcal{H}) &=2(y_{(1,2),\emptyset} \cdot 0  + y_{(1,2),\{1,2\}} \cdot 0) + c_g(y_{g,\{s,1\}}\cdot w_2 + y_{g,\{s,2\}} \cdot w_2)
	= w_2. 
\end{align*}
The gap between the integral and fractional solution is therefore
\begin{equation*}
	\frac{\mathit{OPT}(\mathcal{H})}{\mathit{LP}(\mathcal{H})} = \frac{1}{w_2}.
\end{equation*}
Thus, the Basic LP integrality gap is at least $\frac{1}{w_2}$, and we have the following corollary of Theorem~\ref{thm:ene}.
\begin{corollary}
	Assuming the Unique Games Conjecture, $\wcbcut{}(4,(w_1 = 1, w_2 < 1))$ cannot be approximated to within a factor better than $1/w_2$.
\end{corollary}
This lower bound matches the approximation we get by projecting to the nearest submodular penalties ($\hat{w}_1 = 1, \hat{w}_2 = 1)$, showing that this simple projection is optimal assuming UGC. We will now see that the same projection-based argument also achieves optimality for the case $w_2>2$. Later in the appendix, we present additional integrality gap instances for some more general, non-submodular scenarios with $r>4$ and show the optimality of the nearest submodular projection by finding the convex hull of $\vw$. 

\paragraph{Integrality gap for $w_2 > 2$}
\begin{figure}  
	\subfigure[Optimal solution with $s$-$t$ value $w_2$]  
	{  
		\label{fig:igap_opt_case2}
		\begin{tikzpicture}[scale=.45]  
			
			\definecolor{lightorange}{RGB}{255,204,153}
			\definecolor{lightyellow}{RGB}{255,255,153}
			\definecolor{gray}{RGB}{200,200,200}
			\definecolor{maroon}{rgb}{0.5, 0, 0} 
			
			\node[circle, fill=lightorange, draw, minimum size=0.5cm] (s) at (1,3) {$s$};
			\node[circle, fill=gray, draw, minimum size=0.5cm] (1) at (4,6) {$1$};
			\node[circle, fill=gray, draw, minimum size=0.5cm] (2) at (4,3) {$2$};
			\node[circle, fill=gray, draw, minimum size=0.5cm] (3) at (1,6) {$3$};
			\node[circle, fill=gray, draw, minimum size=0.5cm] (4) at (7,6) {$4$};
			\node[circle, fill=lightorange, draw, minimum size=0.5cm] (t) at (7,3) {$t$};
			
			\draw[thick,maroon] (s) -- (3);
			\draw[thick,maroon] (1) -- (2);
			\draw[thick,maroon] (4) -- (t);
			
			\draw[rounded corners=9pt,thick,maroon] (-1,2) rectangle (5,7);  
			\draw[rounded corners=9pt, thick,maroon] (2.5,1.5) rectangle (8.0,7.5);  
			\draw [thick, dashed] (2,1) to[line to] (2,8); 
			
			\draw[maroon] (-1.5,4.5) node {$g$}; 
			\draw[maroon] (-1.0,1.5) node {$c_g=1$};
			\draw[maroon] (0.2,4.5) node {$2w_2$};  
			\draw[maroon] (3.3,4.5) node {$2w_2$}; 
			\draw[maroon] (6.2,4.5) node {$2w_2$}; 
			\draw[maroon] (8.5,4.5) node {$f$}; 
			\draw[maroon] (9.2,3.5) node {$c_f=1$};
			
			\draw (0.5,0.0) node {$x_{s,s}=1$};
			\draw (0.2,7.5) node {$x_{3,s}=1$};
			\draw (4.0,8.2) node {$x_{1,s}=0$};
			\draw (4.5,0.2) node {$x_{2,s}=0$};
			\draw (8.0,8.0) node {$x_{4,s}=0$};
			\draw (8.0,1.0) node {$x_{t,s}=0$};
		\end{tikzpicture}  
	} \hfill
	\subfigure[Basic LP solution with objective value $2$] { 
		\label{fig:igap_blp_case2}
		\begin{tikzpicture}[scale = 0.45]
			\definecolor{lightorange}{RGB}{255,204,153}
			\definecolor{lightyellow}{RGB}{255,255,153}
			\definecolor{gray}{RGB}{200,200,200}
			\definecolor{maroon}{rgb}{0.5, 0, 0} 
			
			\node[circle, fill=lightorange, draw, minimum size=0.5cm] (s) at (1,3) {$s$};
			\node[circle, fill=gray, draw, minimum size=0.5cm] (1) at (4,6) {$1$};
			\node[circle, fill=gray, draw, minimum size=0.5cm] (2) at (4,3) {$2$};
			\node[circle, fill=gray, draw, minimum size=0.5cm] (3) at (1,6) {$3$};
			\node[circle, fill=gray, draw, minimum size=0.5cm] (4) at (7,6) {$4$};
			\node[circle, fill=lightorange, draw, minimum size=0.5cm] (t) at (7,3) {$t$};
			
			\draw[thick,maroon] (s) -- (3);
			\draw[thick,maroon] (1) -- (2);
			\draw[thick,maroon] (4) -- (t);
			
			\draw[rounded corners=9pt,thick,maroon] (-1,2) rectangle (5,7);  
			\draw[rounded corners=9pt, thick,maroon] (2.5,1.5) rectangle (8.0,7.5);  
			
			\draw[maroon] (-1.5,4.5) node {$g$}; 
			\draw[maroon] (-1.0,1.5) node {$c_g=1$};
			\draw[maroon] (0.2,4.5) node {$2w_2$};  
			\draw[maroon] (3.3,4.5) node {$2w_2$}; 
			\draw[maroon] (6.2,4.5) node {$2w_2$}; 
			\draw[maroon] (8.5,4.5) node {$f$}; 
			\draw[maroon] (9.2,3.5) node {$c_f=1$};
			
			\draw (0.4,0.0) node {$x_{s,s}=1$};
			\draw (0.2,7.5) node {$x_{3,s}=1$};
			\draw (4.0,8.3) node {$x_{1,s}=0.5$};
			\draw (4.5,0.2) node {$x_{2,s}=0.5$};
			\draw (8.0,8.0) node {$x_{4,s}=0$};
			\draw (8.0,1.0) node {$x_{t,s}=0$};
			
	\end{tikzpicture}  }
	
	\caption{Instance for $\ecbcut(4,(w_1=1,w_2>2w_1))$}
	\label{fig:igap_case2}
\end{figure}

Consider an instance of $\ecbcut(4,(w_1 = 1, w_2 >2))$ given by the hypergraph $\mathcal{H} = (\V,\E \cup E)$ with six nodes $\V = \{1,2,3,4,s,t\}$, three edges, and two hyperedges as shown in Figure~\ref{fig:igap_case2}. The hyperedges $f,g \in \E$ have weight $c_g = c_f = 1$, and edges $(s,3)$, $(t,4)$, and $(1,2)$ have weight $2w_2$. 
The minimum $s$-$t$ cut solution has a cut value of $\mathit{OPT}(\mathcal{H}) = w_2$, which can be achieved by cutting the hyperedge $g$ or $f$ in an even $(2,2)$ split. One such division with cut set $S = \{s,3\}$ is illustrated in Figure \ref{fig:igap_opt_case2}, along with binary feasible variables for this solution. The Basic LP has a feasible solution where $x_{3,s} = x_{4,t} = 1$, $x_{4,s} = x_{3,t} = 0$, and $x_{1,s} = x_{1,t} = x_{2,s} = x_{2,t} = 0.5$. 

For $g = (s,1,2,3)$, the edge variables are $y_{g,\{s,2,3\}} = y_{g,\{s,1,3\}} = 0.5$ and $y_{g,A} = 0$ for every $A\notin \{\{s,2,3\},\{s,1,3\}\}$. 
We can confirm that the constraints in~\eqref{lp1} and~\eqref{lp2} are satisfied since
\begin{align*}
	\begin{array}{ll}
		x_{1,s} = y_{g,\{s,1,3\}} = 0.5 &\quad x_{1,t} = y_{g,\{s,2,3\}}= 0.5\\
		x_{2,s} = y_{g,\{s,2,3\}} = 0.5 &\quad x_{2,t} = y_{g,\{s,1,3\}} = 0.5\\
		x_{3,s} = y_{g,\{s,1,3\}} + y_{g,\{s,2,3\}} = 1 &\quad x_{3,t} = 0 \\
		x_{s,s} = y_{g,\{s,1,3\}} + y_{g,\{s,2,3\}} = 1 & \quad x_{s,t} = 0.
	\end{array}
\end{align*}
For $f = (t,1,2,4)$, set $ y_{f,\{2\}} = y_{t,\{1\}} = 0.5$ and $y_{f,A} = 0$ for $A \notin \{\{2\},\{1\}\}$. We see that the constraints are satisfied due
\begin{align*}
	\begin{array}{ll}
		x_{1,s} = y_{f,\{1\}} = 0.5 &\quad x_{1,t} = y_{f,\{2\}} = 0.5\\
		x_{2,s} = y_{f,\{2\}} = 0.5 &\quad x_{2,t} =  y_{f,\{1\}} = 0.5\\
		x_{4,s} = 0  &\quad x_{4,t} = y_{f,\{2\}} + y_{f,\{1\}} = 1\\
		x_{t,s} = 0 &\quad x_{t,t} =  y_{f,\{2\}} + y_{f,\{1\}}=1.
	\end{array}
\end{align*}
For edges $(s,3)$ and $(t,4)$, set $y_{(s,3),\{s,3\}} = 1$ and $y_{(t,4),\{\emptyset\}} = 1$. These also satisfy constraints since $x_{3,s} = 1$ and $x_{4,t} = 1$. For $e = (1,2)$, set $y_{e,\{1,2\}} = y_{e,\emptyset} = 0.5$, satisfying constraints
\begin{align*}
	\begin{array}{ll}
		x_{1,s} = y_{e,\{1,2\}} = 0.5 &\quad x_{1,t} = y_{e,\emptyset} = 0.5\\
		x_{2,s} = y_{e,\{1,2\}} = 0.5 &\quad x_{2,t} = 0.5.
	\end{array}
\end{align*}
The LP value associated with this feasible solution is given by
\begin{align*}
	\mathit{LP}(\mathcal{H}) &= c_g \left[y_{g,\{s,2,3\}}  \vw_g(\{s,2,3\}) + y_{g,\{s,1,3\}} \vw_g(\{s,1,3\})\right]\\ 
	& \hspace{10pt}+ c_f\left[y_{f,\{1\}} \vw_f(\{1\}) + y_{f,\{2\}}  \vw_f(\{2\}) \right]\\
	&=c_g \cdot w_1 + c_f \cdot w_1 = 2.
\end{align*}
Consequently, the integrality gap is $\mathit{OPT}(\mathcal{H})/\mathit{LP}(\mathcal{H}) = w_2 /2$.
\begin{corollary}
	Assuming the Unique Games Conjecture, $\wcbcut{}(4,(w_1 = 1, w_2 >2))$ cannot be approximated to within a factor better than $w_2/2$.
\end{corollary}
The appendix includes other integrality gap instances that prove UGC-hardness of approximation results for certain classes of non-submodular functions when $r > 4$.

\section{Conclusion}
This paper provides two approaches for settling a recently-posed question on the tractability of non-submodular cardinality-based hypergraph $s$-$t$ cut problems~\citep{aksoy2023seven}. Our results confirm that the latter problem is NP-hard for all non-submodular parameter choices except for a degenerate case where a zero-cost solution is easy to achieve. After settling this open question, we show several refined approximation hardness results and a projection-based approximation algorithm. We prove that the latter algorithm gives the best possible approximation for all 4-uniform problems assuming the Unique Games Conjecture. We strongly conjecture that this projection technique is the best possible (assuming UGC) for hypergraphs with arbitrary hyperedge sizes; establishing a proof of this is an open direction for future work. Another open direction is to further tighten the hardness of approximation results that depend only on the assumption that $\mathit{P} \neq \mathit{NP}$. Finally, a more applied direction is to use our approximation techniques to improve downstream hypergraph clustering problems where the most meaningful choice of cut function is only approximately submodular. 

\bmhead{Acknowledgements}
Nate Veldt and Vedangi Bengali are supported by the Army Research Office (ARO award \#W911NF‐24-1-0156). Iiro Kumpulainen and Nikolaj Tatti are supported by the Academy of Finland project MALSOME (343045). We thank Jon Kleinberg and Magnus Wahlstr{\"o}m for several helpful conversations.

\bibliography{references}

\begin{appendices}
	
	\section{UGC hardness results for $r>4$}\label{secA1}
	In this section, we show UGC hardness results for certain cases of non-submodular splitting functions $\vw$ when $r>4$. For these cases, we construct hypergraph instances where the Basic LP exhibits an integrality gap that exactly matches the approximation bound $\rho$ we obtain by projecting $\vw$ onto the nearest submodular $\hat{\vw}$. This projection is obtained by finding the convex hull for the function $\vw$, as detailed in 
	Section~\ref{subsec:convexhull}.
	
	\begin{lemma}
		\label{lem:nphardnessr}
		Assuming Unique Games Conjecture, $\wcbcut(r, \vw)$ for a given $r>4$ cannot be approximated to within a factor better than $\frac{w_i}{w_j}$ when the non-submodular cut penalties $\{w_1=1,w_2,\ldots,w_q\}$ (where $q = \floor{\frac{r}{2}}$) satisfy:
		\begin{align}\label{eq:ugc-inequality1}
			&w_i > w_j \quad \text{ for some } i < j.
		\end{align}
	\end{lemma}
	\begin{proof}
		Given an $\ecbcut(r,\vw)$ problem where $\vw$ satisfies the above inequalities, we call each pair $(i,j)$ a ``bad" pair when $w_i>w_j$ and $i<j$. 
		Now for each bad pair $(i,j)$, we construct a hypergraph instance $\mathcal{H} = (\V,\E \cup E)$ 
		where $\V = \{s_1,\ldots,s_i,u_1,\ldots,u_k, v_1,\ldots,v_{k'},t_1,\ldots,t_i\}$ is a set of $r$ nodes with $k = j-i$ and $k' = r-j-i$.
		Here $s = s_1$ is the source node and is part of the clique formed by the set of nodes $\{s_1,\ldots,s_i\}$. We call this clique $C_s$.  Similarly, with $t = t_1$ as the sink node, nodes in set $\{t_1,\ldots,t_i\}$ form a clique which we refer to as $C_t$. In addition, we have a clique $C_v$ of $k=(j-i)$ nodes $\{v_1,\ldots,v_k\}$, and another clique $C_u$ of $k' = (r-j-i)$ nodes $\{u_1,\ldots,u_{k'}\}$. Every edge within the cliques $C_s,C_t,C_u,C_v$ has a weight of $2w_i$. Nodes $u_1$ and $v_1$ are connected by an edge of weight $c_{(u_1,v_1)} = 2w_i$ and a single hyperedge $e\in \E$ of unit weight includes all of the $r$ nodes. 
		
		\begin{figure}[]
			\centering
			
			\begin{tikzpicture}[scale=0.7]
				\tikzstyle{vertex}=[circle, fill=gray, draw, minimum size=0.75cm, inner sep=0pt]
				
				
				\definecolor{lightorange}{RGB}{255,204,153}
				\definecolor{orange}{RGB}{255,153,51}
				\definecolor{red}{RGB}{255,102,102}
				\definecolor{lightyellow}{RGB}{255,255,153}
				\definecolor{gray}{RGB}{200,200,200}
				\definecolor{maroon}{rgb}{0.5, 0, 0} 
				
				\draw[thick] (0,0) ellipse (9cm and 3.5cm);
				
				\draw[red, thick,dashed] (3.0,-4) -- (3.0,4);

				\node[vertex, fill=lightorange] (s) at (-5,0) {$s_1$};
				\node[vertex] (s2) at (-6.5,1) {$s_3$};
				\node[vertex] (sk) at (-3.5,1) {$s_{i}$};
				\node[vertex] (s1) at (-5,-1.5) {$s_2$};
				
				\draw (s) -- (s1);
				\draw (s) -- (s2);
				\draw (s) -- (sk);
				\draw (s1) -- (s2);
				\draw[thick,dotted] (s2) -- (sk);
				\draw (s1) -- (sk);
				
				\node[vertex, fill=lightorange] (t) at (5.5,0) {$t_1$};
				\node[vertex] (t2) at (4,1) {$t_3$};
				\node[vertex] (tk) at (7,1) {$t_{i}$};
				\node[vertex] (t1) at (5.5,-1.5) {$t_2$};
				
				\draw (t) -- (t1);
				\draw (t) -- (t2);
				\draw (t) -- (tk);
				\draw (t1) -- (t2);
				\draw [thick,dotted] (t2) -- (tk);
				\draw (t1) -- (tk);
				
				\node[vertex] (x) at (0,-1) {$u_1$};
				\node[vertex] (x1) at (0,-2.5) {$u_2$};
				\node[vertex] (x2) at (1.5,-1.5) {$u_{k'}$};
				\node[vertex] (y) at (0,1) {$v_1$};
				\node[vertex] (y1) at (-1.5,1.5) {$v_2$};
				\node[vertex] (y2) at (0,2.5) {$v_k$};
				
				\draw (x) -- (y);
				\draw (x) -- (x1);
				\draw (x) -- (x2);
				\draw[thick,dotted] (x1) -- (x2);
				\draw (y) -- (y1);
				\draw (y) -- (y2);
				\draw[thick,dotted] (y1) -- (y2);
				
				\node at (3.0,-4.5) {$w_{i}$};
				\node at (-6.3,-0.5) {$2w_{i}$};
				\node at (7.0,-0.3) {$2w_{i}$};
				\node at (-0.5,0) {$ 2w_i$};
				\node at (-2.5,-2.9) {$c_{e} = 1$};
				\node at (-1.5,-3.7) {${e}$};
				\node at (-1.0,2.5) {$ 2w_i$};
				\node at (1.0,-2.5) {$ 2w_i$};
				
			\end{tikzpicture}
			\caption{Integrality gap instance for $\ecbcut(r,\vw)$ when $w_i>w_j$ where $i<j$. Here $k = i-j$ and $k' = r-i-j$. $C_s$ is the left clique of $\{s_1,\ldots,s_i\}$ nodes and $C_t$ is the right clique of $\{t_1,\ldots,t_i\}$ nodes. Here the source node is $s_1$ and the sink node is $t_1$. $C_u$ and $C_v$ form the middle two cliques of $\{u_1,\ldots,u_{k'}\}$  and $\{v_1,\ldots,v_k\}$ vertices respectively.}
			\label{fig:ugc-first}
		\end{figure}
		
		The minimum $s$-$t$ cut solution cuts hyperedge $e$ in an $(i,r-i)$ split with $OPT(\mathcal{H}) = w_i$. This is obtained by cutting $e$ such that either the clique $C_s$ or $C_t$ is placed in its own partition. One such cut is shown in Figure~\ref{fig:ugc-first}.
		The Basic LP gives a fractional solution where $x_{z,s} = x_{z,t} = 0.5$ for all nodes $z \in \{u_1,\ldots,u_{k'},v_1,\ldots,v_{k}\}$, and $x_{s',s} = x_{t',t} = 1$ for $s' \in \{s_1,\ldots,s_i\}$ and 
		$t' \in \{t_1,\ldots,t_i\}$. In what follows, we show how to set edge variables $y_{f,A}$ for every $f \in \E \cup E$ and every $A \subseteq f$ so that constraints~\eqref{lp1} and~\eqref{lp2} are satisfied. 
		
		Write $A_1 = \{ s_1, \ldots, s_i, u_1, v_1, \ldots, v_{k-1} \}$ and $A_2 = \{ s_1, \ldots, s_i, v_k, u_2, \ldots, u_{k'} \}$.
		For the hyperedge $e$, we set
		\[
		y_{e,A_1} = y_{e,A_2} = 0.5
		\]
		and $y_{e,A} = 0$ for all remaining subsets $ A \neq A_1, A_2$. We can confirm for every node in $e$, conditions~\eqref{lp1} and~\eqref{lp2} hold since 
		\begin{align*}
			x_{z,s} &= \sum_{A \subseteq e:z \in A} y_{e,A}= 0.5, && \text{for } z \in C_u,C_v,\\
			x_{z,t} &= \sum_{A \subseteq e:z \in e\setminus A} y_{e,A}= 0.5, && \text{for } z \in C_u,C_v,\\
			x_{z,s} &= y_{e,A_1} + y_{e,A_2} = 1.0, && \text{for } z\in C_s,\\
			x_{z,t} &= y_{e,A_1} + y_{e,A_2} = 1.0, & & \text{for } z\in C_t.
		\end{align*}

		For every edge $f$ within cliques $C_u$ and $C_v$, we set $y_{f,\emptyset} = y_{f,f} = 0.5$, and $y_{f,A} = 0$ for $A \notin \{\emptyset,f\}$. For each of these edges $f$, constraints~\eqref{lp1} and~\eqref{lp2} are satisfied since
		\begin{align*}
			x_{z,s} = y_{f,f} = 0.5 \quad\text{and}\quad x_{z,t} = y_{f,\emptyset} = 0.5, \quad\text{where}\quad z \in f.
		\end{align*}
		
		On the other hand, variable $y_{f,A}$ for every edge within the clique $C_s$ is set to $1$ for $A = f$, and set to $0$ for $A \neq f$. Similarly for all edges in $C_t$, we set $y_{f,\emptyset} = 1$ and $y_{f,A} = 0$ for $A \neq \emptyset$. We can confirm for each node in $C_s$ and $C_t$, conditions~\eqref{lp1} and~\eqref{lp2} hold since
		\begin{align*}
			&\text{ for each } z \in C_s \text{ and adjacent edge } f, \quad x_{z,s} = y_{f,f} = 1 \text{ and } x_{z,t} = y_{f,\emptyset} = 0, \text{ and }\\
			&\text{ for each } z \in C_t \text{ and adjacent edge } f, \quad x_{z,t} = y_{f,\emptyset} = 1 \text{ and } x_{z,s} = y_{f,f} = 0. 
		\end{align*} 
		Finally, for the edge $(u_1,v_1)$, we have $y_{(u_1,v_1),A} = 0.5 $ for $A \in \{\emptyset,\{u_1,v_1\}\}$ and $0$ for $A \notin \{\emptyset,\{u_1,v_1\}\}$. Thus,
		\begin{align*}
			&x_{u_1,s} = x_{v_1,s} = y_{(u_1,v_1),\{u_1,v_1\}} = 0.5 \text{ and } \\
			&x_{u_1,t} = x_{v_1,t} = y_{(u_1,v_1),\emptyset} = 0.5.\\
		\end{align*}
		The LP value for this feasible solution is
		\begin{align*}
			LP(\mathcal{H}) &= c_e(y_{e,A_1}\cdot \vw_e(A_1) + y_{e,A_2}\cdot \vw_e(A_2) )
			+ 2w_i(y_{(u_1,v_1),\{u_1,v_1\}} \cdot 0+ y_{(u_1,v_1),\emptyset} \cdot 0) \\
			&=c_e(0.5 \cdot w_j + 0.5 \cdot w_j ) = w_j. 
		\end{align*}
		This gives us an integrality gap of 
		\begin{align}
			\frac{OPT(\mathcal{H})}{LP(\mathcal{H})} = \frac{w_i}{w_j},
		\end{align}
		proving the claim.
	\end{proof}
	
	We can thus get an integrality gap of $\frac{w_i}{w_j}$ for every bad pair $(i,j)$ in a given set of penalties $\{w_1,\ldots,w_q\}$ by constructing its corresponding hypergraph instance. 
	This proves that it is UGC-hard to approximate \wcbcut{}($r,\vw$) with $\vw$ satisfying inequalities in Lemma~\ref{lem:nphardnessr} to a factor better than the maximum ratio $\frac{w_i}{w_j}$ over all the bad pairs $(i,j)$. This factor matches the approximation bound obtained by finding the largest gap between the discrete values of convex hull $h(i)$ and $\vw(i)$ in certain cases, discussed in the following corollary.

	\begin{corollary}\label{cor:ugc1convexhull}
		Consider a non-submodular $\wcbcut{}(r,\vw)$ problem where for some integer $t \in \{1,2, \ldots, q\}$ the splitting penalties satisfy:
		\begin{align*}
			w_{i} &\leq w_{i+1} & \text{ for $i \leq t-1$}\\
			2w_i &\geq w_{i-1} + w_{i+1} &\text{ for $i \leq t-1$} \\
			w_j &< w_t & \text{ if $ t < j \leq q$}.
		\end{align*}
		Let $h$ be the upper non-decreasing convex hull for $\vw$ as defined in Theorem~\ref{thm:convexhull}. The ratio $h(t)/w_t$ exactly matches the Basic LP integrality gap from Lemma~\ref{lem:nphardnessr}.
	\end{corollary}
	\begin{proof}
		Since $h$ is non-decreasing and concave, observe that $h(j) = \hat{w}(j) = w_i$ for all $j \geq t$. The largest gap between $h$ and $\vw$ within the interval $[t,q]$ is then given by the maximum ratio $\frac{h(j)}{\vw(j)} = \frac{w_t}{w_j} $ over all $j \in \{t,q\}$. The approximation bound obtained by projecting to the upper convex hull is therefore optimal assuming UGC.
	\end{proof}
	We now show an integrality gap instance for non-submodular $\vw$ when $2w_i < w_{i-1} + w_{i+1}$ for any $i \in \{1,2,\ldots,q\}$. Here when $i=1$, $w_{i-1} = w_0 = 0$. 
	\begin{lemma}\label{lem:nphardnessr2}
		Approximating \wcbcut{}($r,\vw$) to a factor better than $\frac{w_{i-1}+w_{i+1}}{2w_i}$ is UGC hard when the cut penalties $\{w_1=1,w_2,\ldots,w_q\}$ of the non-submodular $\vw$ satisfy:
		\begin{align*}
			&2w_i < w_{i-1} + w_{i+1} \text{ for any } i \in \{1,2,\ldots,q-1\}.
		\end{align*}
	\end{lemma}
	\begin{proof}
		In this case, we consider a triplet of breakpoints $(i-1,i,i+1)$ as a ``bad" triplet if the corresponding splitting penalties satisfy the above inequality. For each such bad triplet we construct an instance $\mathcal{H} = (\V,\E \cup E)$ of $\ecbcut(r,\vw)$ as shown in Figure~\ref{fig:ugc-second}. Here, $\V$ is the set of $2r - 2i$ nodes $\{s_1,\ldots,s_{r-i-1},u,v,t_1,\ldots,t_{r-i-1}\}$ with $s = s_1$ as the source node and $t = t_1$ as the sink. We first construct a clique of all nodes  in $\{s_1,\ldots,s_{r-i-1}\}$ and call it $C_s$, where each of its edges $(s_j,s_k)$ has a large weight of $c' = 2(w_{i-1}+w_{i+1})$. Similarly we form a clique $C_t$ of all nodes in $\{t_1,\ldots,t_{r-i-1}\}$, and add an edge $(u,v)$ with a weight $c'$. We then add hyperedges $e = (s_1,\ldots,s_{i-1},u,v,t_{1},\ldots,t_{r-i-1}) $ and $f = (s_1,\ldots,s_{r-i-1},u,v,t_1,\ldots,t_{i-1})$; when $i=1$, edge $e$ only consists of nodes $\{u,v,t_1,\ldots,t_{r-2}\}$ and $f = (s_1,\ldots,s_{r-2},u,v)$.
		Both $e,f \in \E$ have weights $c_e = c_f = 1$.
		
		\begin{figure}[]
			\centering
			\begin{tikzpicture}[scale=0.65]
				\tikzstyle{vertex}=[circle, fill=gray, draw, minimum size=0.75cm, inner sep=0pt]

				\definecolor{lightorange}{RGB}{255,204,153}
				\definecolor{lightyellow}{RGB}{255,255,153}
				\definecolor{gray}{RGB}{200,200,200}
				\definecolor{maroon}{rgb}{0.5, 0, 0}

				\draw[maroon] plot[smooth cycle, tension=.7] coordinates {(-8.5,-0.5) (-8,3.5) (-1,4.0) (5,4.0) (8.2,3.2)   (8.2,-4) (4,-4)  (-0.5,-1.5) };
				
				\draw[blue] plot[smooth cycle, tension=.7] coordinates {(9.0,-1.0) (7,5) (0,5) (-7,4) (-7.0 ,-4) (-3,-4) (0,-2) };

				\draw[red, thick,dashed] (-2.5,-5) -- (-2.5,6);

				\node[vertex, fill=lightorange] (s1) at (-5,0) {$s_1$};
				\node[vertex] (s3) at (-6,3) {$s_3$};
				\node[vertex] (si1) at (-4,3) {$s_{i-1}$};
				\node[vertex] (s2) at (-7,0.5) {$s_2$};
				\node[vertex] (sj) at (-3.5,-3) {$s_{j}$};
				\node[vertex] (sr) at (-6,-3) {$s_{r-j}$};
				
				\draw (s1) -- (s2);
				\draw (s1) -- (s3);
				\draw (s1) -- (si1);
				\draw (s2) -- (s3);
				\draw (s2) -- (si1);
				\draw[thick,dotted] (s3) -- (si1);
				\draw (s1) -- (si1);
				\draw (s1) -- (sj);
				\draw (s2) -- (sj);
				\draw (s1) -- (sr);
				\draw (s2) -- (sr);
				\draw (s3) -- (sr);
				\draw[thick,dotted] (sj) -- (si1);
				\draw[thick,dotted] (sj) -- (sr);
				
				\node[vertex, fill=lightorange] (t1) at (6,0) {$t_1$};
				\node[vertex] (t3) at (4,3) {$t_3$};
				\node[vertex] (ti) at (7,3) {$t_{i-1}$};
				\node[vertex] (t2) at (3.5,0.5) {$t_2$};
				\node[vertex] (tr) at (5,-3) {$t_{r-j}$};
				\node[vertex] (tj) at (7.5,-3) {$t_{j}$};
				
				\draw (t1) -- (t2);
				\draw (t1) -- (t3);
				\draw (t2) -- (t3);
				\draw (t1) -- (ti);
				\draw (t2) -- (tj);
				\draw [thick,dotted] (t3) -- (ti);
				\draw [thick,dotted] (ti) -- (tj);
				\draw (t2) -- (ti);
				\draw (t1) -- (tj);
				\draw (t1) -- (tr);
				\draw (t2) -- (tr);
				\draw [thick,dotted](tj) -- (tr);
				\draw (t3) -- (tr);
				
				\node[vertex] (x) at (0,1) {$v$};
				\node[vertex] (y) at (0,3) {$u$};

				\draw (x) -- (y);

				\node at (-1.0,5.5) {$w_{i-1} + w_{i+1}$};
				\node at (-7.0,2) {$c'$};
				\node at (3.2,2) {$c'$};
				\node at (-0.5,2) {$c'$};
				\node at (-3.5,-4.7) {\textcolor{blue}{$c_{f} = 1$}};
				\node at (-5.0,-4.3) {\textcolor{blue}{$f$}};
				\node at (3.8,-4.5) {\textcolor{maroon}{$c_{e} = 1$}};
				\node at (3.0,-3.2) {\textcolor{maroon}{$e$}};
				
			\end{tikzpicture}
			\caption{
				Integrality gap instance for $\ecbcut(r,\vw)$ when $w_{i-1} + w_{i+1} > 2w_i$. 
				Here $c' = 2(w_{i-1} + w_{i+1})$. For aesthetic reasons, we write $s_j$ 
				for $s_{i+1}$ and $s_{r-j}$ for $s_{r-i-1}$ where $j=i+1$. $C_s$ is the left clique 
				of $\{s_1, \ldots, s_{r-i-1}\}$ nodes, and $C_t$ is the right clique 
				of $\{t_1, \ldots, t_{r-i-1}\}$ nodes. Here, the source node is $s_1$, 
				and the sink node is $t_1$. Hyperedge $e$ is shown in maroon and $f$ in blue for better visualization. The red dashed line depicts the optimal cut with a cut value of $w_{i-1}+w_{i+1}$. }
			\label{fig:ugc-second}
		\end{figure}

		The minimum $s$-$t$ cut solution has an optimal cut value of $OPT(\mathcal{H}) = w_{i-1}+w_{i+1}$. One way of obtaining this value is by splitting $e$ with cut value of $w_{i-1}$ and $f$ with a value of $w_{i+1}$ where the cut set is $S = \{s_1,\ldots,s_{r-i-1}\}$ as shown in Figure~\ref{fig:ugc-second}. The binary variables for this feasible solution is $x_{s,s_j} = 1$ and $x_{t,u} = x_{t,v} = x_{t,t_j} = 1$ for all $j \in \{1,\ldots,r-i-1\}$. 
		
		The Basic LP on the other hand gives a feasible solution with $x_{s,s_j} = 1, x_{t,t_j} = 1$ for all $j \in \{1,\ldots,r-i-1\}$ and $x_{s,u} = x_{t,u} = x_{s,v} = s_{t,v} = 0.5$.
		We now confirm that conditions~\eqref{lp1} and~\eqref{lp2}
		hold for every edge in $\E \cup E$.  
		
		Let us define $A_1 = \{s_1,\ldots,s_{i-1},u\}$ and $A_2 = \{s_1,\ldots,s_{i-1},v\}$.
		The edge variables for the edge $e$ are $y_{e,A_1} = y_{e,A_2} = 0.5 $, and $y_{e,A} = 0$ for all $A \neq A_1, A_2$. The fractional variables for nodes in $e$ satisfy  constraints~\eqref{lp1} and~\eqref{lp2} since
		\begin{align*}
			x_{s_j,s} & = y_{e,A_1} + y_{e,A_2} = 1 \quad\text{and}\quad x_{s_j,t} = 0, &&  \text{for all } j \in 1,\ldots,i-1,  \\
			x_{t_k,t} & = y_{e,A_1} + y_{e,A_2} = 1\quad\text{and}\quad x_{t_k,s} = 0,  &&   \text{for all } k \in 1,\ldots,r-i-1,\\
			x_{u,s} & = y_{e,A_1} = 0.5 \quad \text{and} \quad  x_{u,t} = y_{e,A_2}= 0.5,\\
			x_{v,s} & = y_{e,A_2} = 0.5 \quad  \text{and} \quad x_{v,t} = y_{e,A_1} = 0.5.
		\end{align*}
		Let us now define $A_3 = \{s_1,\ldots,s_{r-i-1},u\}$ and $A_4 = \{s_1,\ldots,s_{r-i-1},v\}$.
		For edge $f$, the edge variables are $y_{f,A_3} = y_{f,A_4} = 0.5$ and $y_{f,A}=0$ for $A \neq A_3, A_4$. The node variables then satisfy
		\begin{align*}
			x_{s_k,s} & = y_{f,A_3} + y_{f,A_4} = 1 \quad \text{and} \quad x_{s_k,t} = 0 && \text{for all } k \in 1,\ldots,r-i-1,  \\
			x_{t_j,t} & = y_{f,A_3} + y_{f,A_4} = 1 \quad \text{and} \quad x_{t_j,s} = 0 && \text{for all } j \in 1,\ldots,i-1,\\
			x_{u,s} & = y_{f,A_3} = 0.5 \quad \text{and} \quad  x_{u,t} = y_{f,A_4}= 0.5,\\
			x_{v,s} & = y_{f,A_4} = 0.5 \quad  \text{and} \quad x_{v,t} = y_{f,A_3} = 0.5.
		\end{align*}
		For every edge $g = (s_j,s_k)$ where $s_k,s_j \in C_s$, set $y_{g,g} = 1$ and similarly for every edge $h = (t_j,t_k)$ where $t_k,t_j \in C_t$, set $y_{h,\emptyset} = 1$. In this case, node variables are $x_{s_j,s} = x_{s_k,s} = y_{g,g} = 1 $ and $x_{t_j,t} = x_{t_k,t} = y_{h,\emptyset} = 1$. Finally for edge $(u,v)$, we have $y_{(u,v),\{u,v\}} = y_{(u,v),\emptyset}  = 0.5$ which satisfy constraints~\eqref{lp1} and~\eqref{lp2} since
		\begin{align*}
			x_{u,s} &= y_{(u,v),\{u,v\}} = 0.5, \quad &x_{u,t} &= y_{(u,v),\emptyset} = 0.5,\\
			x_{v,s} &= y_{(u,v),\{u,v\}} = 0.5, \quad &x_{v,t} &= y_{(u,v),\emptyset} = 0.5.
		\end{align*}
		The LP value of this fractional feasible solution is given by 
		\begin{align*}
			LP(\mathcal{H}) &= c_{e}[y_{e,A_1}\vw_e(A_1) + y_{e, A_2}\vw_e(A_2)] 
			+ c_f[y_{f,A_3 }\vw_f(A_3) + y_{f, A_4}\vw_f(A_4)]\\
			&= c_e\cdot w_i + c_f\cdot w_i = 2w_i.
		\end{align*}
		Hence the integrality gap is $\frac{OPT(\mathcal{H})}{LP(H)} = \frac{w_{i-1} + w_{i+1}}{2w_i}.$
	\end{proof}
	
	Under certain conditions, we can achieve an approximation bound that matches this integrality gap by finding the upper non-decreasing convex hull $h$ for projecting $\vw$. The following corollary identifies a class of non-submodular instances (as characterized in Lemma~\ref{lem:nphardnessr2}) for which this nearest projection strategy is optimal. 
	
	\begin{corollary}\label{cor:ugc2convexhull}
		Consider a non-submodular $\wcbcut{}(r,\vw)$ problem where $w_i \leq w_{i+1}$ for every $i \in \{1,2, \hdots, q\}$ and for specific integer $t \in \{1,2,\ldots,q-1\}$ satisfies
		\begin{align}
			2w_i < w_{i-1}+w_{i+1} & \text{ if $i = t$}\\
			2w_i \geq w_{i-1}+w_{i+1} & \text{ if $i \in \{1,2, \hdots, q\}\setminus \{t\}$}
		\end{align}
		Let $h$ be the upper non-decreasing convex hull for $\vw$ as defined in Theorem~\ref{thm:convexhull}. The ratio $h(t)/w_t$ exactly matches the Basic LP integrality gap from Lemma~\ref{lem:nphardnessr2}.
		\begin{proof}
			Given a non-submodular $\vw$ that meets the above conditions, focus on the interval $[t-1,t+1]$. In this interval, $h$ forms a linear segment connecting $(t-1,w_{t-1})$ and $(t+1,w_{t+1})$ with a midpoint of $h(t) = (w_{t-1}+w_{t+1})/{2}$. Since $2w_t < w_{t-1}+w_{t+1}$, we have a gap $\frac{h(t)}{\vw(t)} = \frac{w_{t-1}+w_{t+1}}{2w_t} $ which equals the integrality gap in Lemma~\ref{lem:nphardnessr2}. 
		\end{proof}
	\end{corollary}

	Although conditions in Corollaries~\ref{cor:ugc1convexhull} and~\ref{cor:ugc2convexhull} do not capture all non-submodular scenarios, we strongly conjecture that, assuming UGC, maximizing the ratio $h(i)/\vw(i)$ over all $i$ in $\{1,2,\ldots,q\}$ yields an optimal approximation factor for any non-submodular $\wcbcut{}(r,\vw)$ problem.
	
\end{appendices}

\end{document}